\documentclass{aa}  
\usepackage{graphicx}
\usepackage{natbib}
\usepackage{txfonts}
\begin{document}
   \title{Testing grain surface chemistry : a survey of deuterated formaldehyde and methanol in low-mass Class 0 protostars\thanks{Based on observations with the IRAM 30\,m telescope at Pico Veleta (Spain). IRAM is funded by the INSU/CNRS (France), the MPG (Germany) and the IGN (Spain).}}

   \subtitle{}

   \author{   B. Parise\inst{1}
  \and C. Ceccarelli\inst{2}  \and A.G.G.M. Tielens\inst{3}  \and A. Castets\inst{4}   \and E. Caux\inst{5}  \and B. Lefloch\inst{2}   \and S. Maret\inst{6} 
                  }
   
   \offprints{B. Parise}

   \institute{Max Planck Institut f\"ur Radioastronomie, Auf dem H\"ugel 69, 53121 Bonn, Germany
\and
Laboratoire d'Astrophysique, Observatoire de Grenoble -BP 53, F-38041
Grenoble cedex 09, France
\and
SRON, P.O. Box 800, NL-9700 AV Groningen, the Netherlands
\and
Observatoire de Bordeaux, BP 89, 33270 Floirac, France
\and
CESR CNRS-UPS, BP 4346, 31028 - Toulouse cedex 04, France
\and 
Department of Astronomy, University of Michigan, 500 Church Street, Ann Arbor MI 48109-1042, USA
        }
  \titlerunning{Deuterated formaldehyde and methanol in low-mass protostars }
   
   \date{Received September 15, 1996; accepted March 16, 1997}

 
  \abstract
   {Despite the low cosmic abundance of deuterium (D/H $\sim$ 10$^{-5}$), large degrees of deuterium fractionation in molecules are observed in star forming regions with enhancements that can reach 13 orders of magnitude, which current models have difficulties to account for.}
   {Multi-isotopologue observations are a very powerful constraint for chemical models. The aim of our  
   observations is to understand the processes forming the observed large abundances of methanol and formaldehyde in low-mass protostellar envelopes (gas-phase processes ? chemistry on the grain surfaces ?) and better constrain the chemical models. }
   {Using the IRAM 30m single-dish telescope, we observed deuterated formaldehyde (HDCO and D$_2$CO) and methanol (CH$_2$DOH, CH$_3$OD, and CHD$_2$OH) towards a sample of seven low-mass class 0 protostars. Using population diagrams, we then derive the fractionation ratios of these species (abundance ratio between the deuterated molecule and its main isotopologue) and compare them to the predictions of grain chemistry models.  }
   {These protostars show a similar level of deuteration as in IRAS16293$-$2422, where doubly-deuterated methanol -- and even triply-deuterated methanol -- were first detected. Our observations point to the formation of methanol on the grain surfaces, while formaldehyde formation cannot be fully pined down. While none of the scenarii can be excluded (gas-phase or grain chemistry formation), they both seem to require abstraction reactions to reproduce the observed fractionations.   }
   {}
   \keywords{ISM: abundances -- ISM:
molecules -- Stars: formation 
               }
   \maketitle
%

\section{Introduction}
In the last few years, observations of low-mass protostars have revealed unexpected high 
abundances of deuterated molecules, and particularly doubly-deuterated 
molecules. The discovery of an extremely large amount (D$_2$CO/H$_2$CO 
$\sim$ 10\%) of doubly-deuterated formaldehyde in the low-mass protostar
IRAS16293$-$2422 \citep[hereafter IRAS16293;][]{Ceccarelli98} was 
followed by observations of this same molecule towards a large sample of 
low-mass protostars \citep{Loinard02}. 
The fractionation ratios appeared to be similarly large towards all 
targeted low-mass protostars. The suggested interpretation was that such large 
deuteration is obtained during the cold and dense precollapse phase 
of the low-mass protostars: highly-deuterated ices are very likely 
formed via active grain surface chemistry \citep{Tielens83}, 
stored in the grain mantles and eventually released in the gas phase 
during the collapse, when the heating of the newly-formed protostar 
evaporates the ices \citep{Ceccarelli01}. A strong support to this scheme comes from the 
large abundance of D$_2$CO observed in prestellar cores \citep{Bacmann03}, 
and from the discovery of the large fractionation of H$_3^+$ in the same objects \citep{Caselli03,Vastel04}. While these last observations undoubtly supported the idea that 
deuteration sets on just before the collapse, the leading chemical process was still largely unproven.
In this context, methanol is a key molecule. Indeed, it is believed to be a grain surface product as
gas-phase processes are not efficient enough to account for the large abundances observed 
in star-forming regions \citep{Herbst05}. Methanol may be the last step of CO hydrogenation on the grain surfaces, after the formaldehyde formation. If formaldehyde formation is also dominated by grain 
chemistry, the deuteration of both molecules is expected to be tightly linked.   

Three years ago, doubly-deuterated methanol was detected towards the 
low-mass protostar IRAS16293 \citep[CHD$_2$OH/CH$_3$OH $\sim$ 20\%,][] 
{Parise02}, basically confirming the grain chemistry scheme. However, 
the CH$_2$DOH/CH$_3$OD 
ratio was found to be unexpectedly large \citep[20 $\pm$ 14,][]{Parise02} 
compared to the value of 3 predicted by the models 
\citep{Charnley97}. The proposed hypothesis of a rapid conversion of CH$_3$OD into CH$_3$OH in the  gas phase due to protonation reactions that would affect only the species for which the deuterium  is bound to the very electronegative oxygen \citep{Charnley97} was also suggested by the model of \citet{Osamura04}. The high observed fractionation of CH$_2$DOH and CHD$_2$OH was consistent with formation of methanol on the grain surfaces, but required an atomic D/H ratio in the gas-phase as high as 0.1$-$0.2 during the mantle formation. This value was challenging gas-phase models at that time, and led \citet{Parise02} to suggest that "a key parameter was missing" in the chemical schemes. Indeed, this key parameter was soon discovered. 
In molecular clouds, the main reservoir of deuterium is molecular HD. Some deuterium can be transferred from this reservoir to other molecules by the intermediate of the ion H$_2$D$^+$ that forms according to the exothermic reaction:
${\rm  HD + H_3^+ \rightarrow H_2 + H_2D^+ }$. Collision of CO with H$_2$D$^+$, followed 
by recombination will then form atomic deuterium. Depletion of CO --- one of the main destruction agents of H$_2$D$^+$ --- drives the fractionation of this key intermediary in the gas phase deuterium fractionation schemes all the way to HD$_2^+$ and even D$_3^+$. 
The inclusion into the 
gas-phase schemes of multiply-deuterated isotopomers of H$_2$D$^+$ as new intermediate molecules for deuterium transfer from the HD main reservoir allowed to predict the 
atomic D/H ratio required by methanol observations \citep{Roberts03}.
Triply-deuterated methanol was later detected in IRAS16293 \citep{Parise04}, and the CH$_3$OH
column density was better evaluated by analyzing $^{13}$CH$_3$OH transitions. The 
observed CD$_3$OH/CH$_3$OH fractionation ratio, that was found to be consistent with the 
CH$_2$DOH and CHD$_2$OH fractionations, allowed to confirm the grain chemistry
scheme. While much progress has been made through laboratory \citep[e.g.][]{Nagaoka05} and theoretical studies to understand the  observed high deuterium fractionation in IRAS16293, observations of a larger sample are required to determine whether this is a common phenomena in protostellar environments.

In this paper, we report singly (HDCO) and doubly (D$_2$CO) deuterated formaldehyde and 
singly-deuterated methanol CH$_2$DOH and CH$_3$OD as well as doubly-deuterated CHD$_2$OH observations towards a sample of low-mass protostars. 
This article is organized as follows: observations are presented in section 2, analysis of the data is presented in section 3, the results are discussed in section 4 and conclusions are drawn in section 5. 


\section{Observations and results}\label{sec-obs}

\subsection{Observations}

Using the IRAM 30-meter telescope (Pico Veleta, Spain), we observed the 
five deuterated species HDCO, D$_2$CO, CH$_2$DOH, CH$_3$OD and CHD$_2$OH 
towards the six low-mass protostars NGC1333$-$IRAS4A, $-$IRAS4B, $-$IRAS2, L1448N, L1448mm and L1157mm. We also present observations of deuterated formaldehyde (HDCO and D$_2$CO) towards L1527. All these sources, already studied by \citet{Maret04},  are Class 0 protostars, {\it i.e.} in the early phase of the gravitational collapse. Table \ref{sources} summarizes the main characteristics of these sources.

\begin{table*}
\caption{Parameters of the sources targeted in our deuterated formaldehyde and methanol observations. IRAS 16293$-$2422 is added for comparison. The offset used for the OFF position of the position switching mode are indicated in the last column.}
\begin{tabular}{ccccccccccc}
\noalign{\smallskip}
\hline
\hline
\noalign{\smallskip}
Source & $\alpha$ (2000) & $\delta$ (2000) & Region & Distance & L$_{\rm bol}$$^a$ & M$_{\rm env}$$^b$ & L$_{\rm smm}$/L$_{\rm bol}$$^a$ & T$_{\rm bol}$ & V$_{LSR}$ & offsets\\
 & & & & \footnotesize (parsec) & \footnotesize (L$_{\odot}$)  & \footnotesize (M$_{\odot}$)  & \footnotesize (\%) & \footnotesize (K) & \footnotesize (km\,s$^{-1}$) & \footnotesize (arcsec) \\
\noalign{\smallskip}
\hline
\noalign{\smallskip}
IRAS4A & 03:29:10.3 & 31:13:32 & Perseus & 220 & 6 & 2.3 & 5 & 34 & +7 & (-150, 240)\\
IRAS4B & 03:29:12.0 & 31:13:09 & Perseus & 220 & 6 & 2.0 & 3 & 36 & +7 & (-180, 260)\\
IRAS2  & 03:28:55.4 & 31:14:35 & Perseus & 220 & 16 & 1.7 & $\leq$ 1 & 50 & +7 & (70, 180)  \\
L1448N   & 03:25:36.3 & 30:45:15 & Perseus & 220 & 6 & 3.5 & 3 & 55 & +5 & (300, 300)\\
L1448mm  & 03:25:38.8 & 30:44:05 & Perseus & 220 & 5 & 0.9 & 2 & 60 & +5 & (300, 300) \\
L1157mm  & 20:39:06.2 & 68:02:22 & Isolated & 325 & 11 & 1.6 & 5 & 60 & +5 & (150, 0) \\
L1527    & 04:39:53.9 & 26:03:10 & Taurus & 140 & 2 & 0.9 & 0.7 & 60 & +5 & (240, 0) \\
\noalign{\smallskip}
\hline
\noalign{\smallskip}
IRAS16293 & 16:32:22.6 & -24:28:33.0 & $\rho$-Ophiucus & 160 & 27 & 5.4 & 2 & 43 & +4 & \\
\noalign{\smallskip}
\hline
\noalign{\smallskip}
\end{tabular}
$^a$ From \citet{Andre00} and \citet{Cernis90}\\
$^b$ From \citet{Jorgensen02}\\
\label{sources}

\end{table*}

The observations were performed in September and November 2002, September 2003 and March 2004. For methanol, four receivers were used simultaneously at 3, 3, 1.3 and 1.3 mm 
with typical system temperatures of about 100, 150, 400 and 400 K 
respectively. These receivers were connected to the VESPA autocorrelator, divided in 6 units.
For formaldehyde, four receivers were used simultaneously at 3, 2, 1.3 and 1 mm, with typical 
system temperatures of about 110, 220, 300, and 400 K, connected to VESPA. 
The telescope beam width varies between 30$''$ at 83 GHz and 9$''$
at 276 GHz. All observations were performed using the position switching 
mode. The offset positions are summarized in Table \ref{sources}.
The pointing accuracy was monitored regularly on strong extragalactic
continuum sources and found to be better than 3$''$. Our spectra for deuterated methanol 
were obtained with integration times ranging from 280 to 420 minutes 
depending on the source.

All fluxes are indicated in units of main-beam temperature, and the error bars were calculated
as the quadratic sum of the statistical noise (noise rms of the data) and the calibration uncertainty.
In order to account for the atmospheric calibration as well as for uncertainties in the band rejection, and taking into account that most of our observations were performed at high elevation,  we adopted the following calibration uncertainties: 5\% for lines observed at frequencies lower than 130\,GHz, 10\% for frequencies between 130 and 260\,GHz, and 15\% for higher frequencies.

All upper limits are given at a 3$\sigma$ level:
$${\rm \int T_{ mb}dv \le 3 \sigma (1+\alpha) \sqrt{\delta v . \Delta v} }$$
where $\alpha$ is the calibration uncertainty indicated above, $\sigma$ the noise rms of the 
observations, $\delta$v the spectral resolution and $\Delta$v the assumed linewidth 
(1.5 km\,s$^{-1}$ for D$_2$CO, 3 km\,s$^{-1}$ for deuterated methanol), based on observations of detected lines. 

\begin{figure*}[!ht]
\includegraphics[width=19cm,angle=-90]{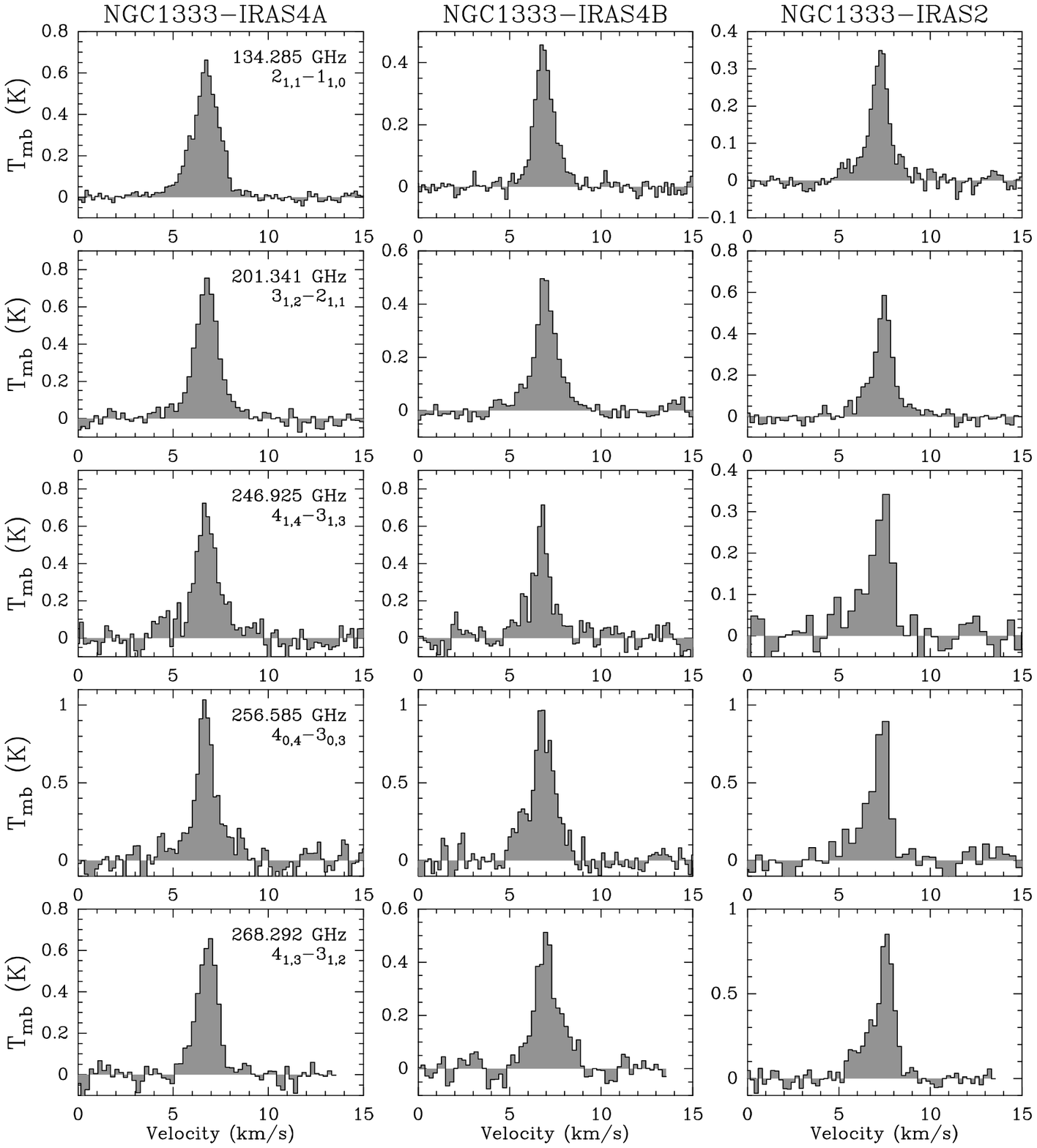}
\caption{HDCO lines for IRAS4a, IRAS4b, IRAS2.}
\label{HDCOlines}
\end{figure*}

\begin{figure*}[!ht]
\includegraphics[width=19cm,angle=-90]{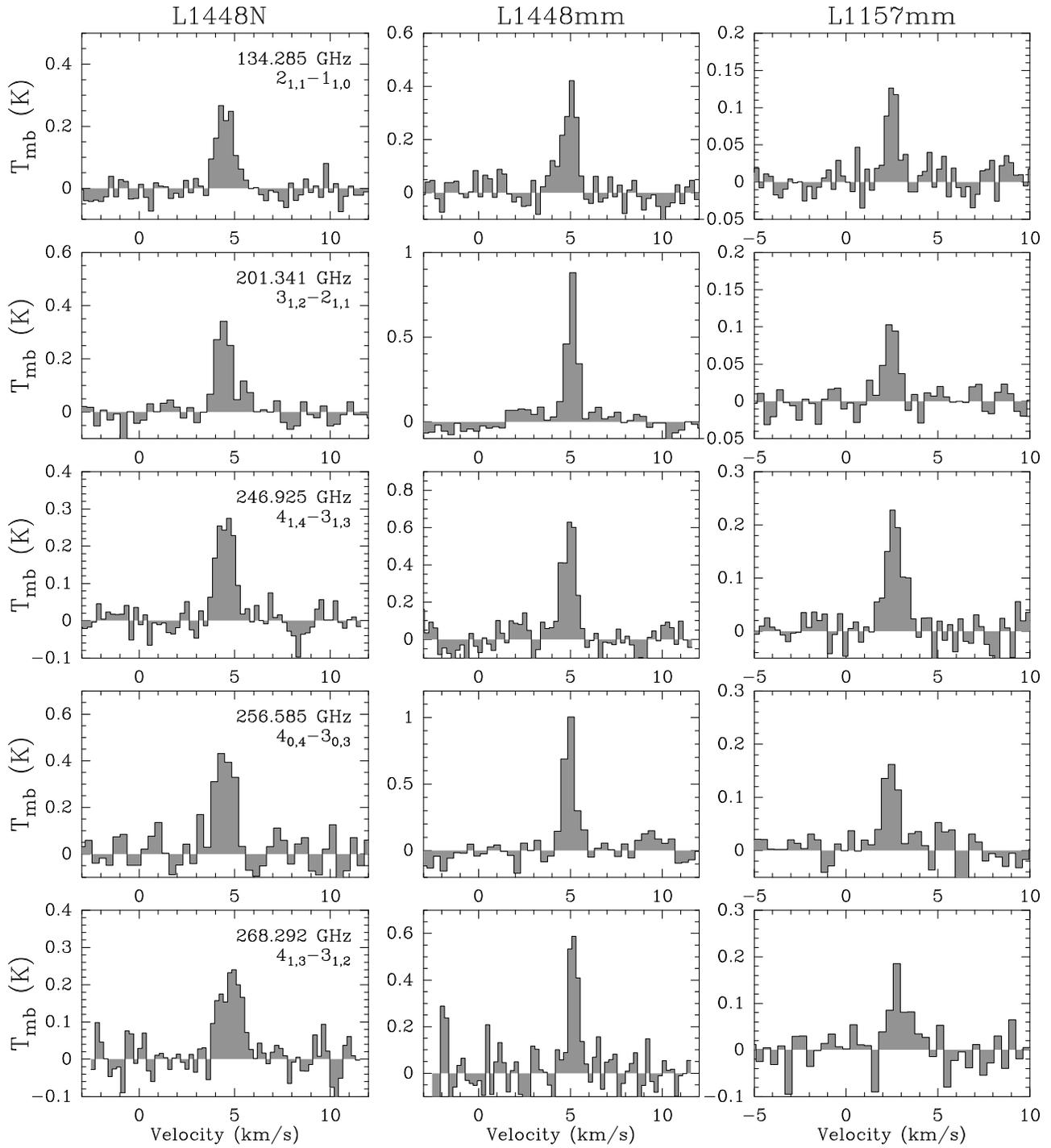}
\caption{HDCO lines for L1448N, L1448mm, L1157mm}
\label{HDCOlines2}
\end{figure*}

\begin{figure*}[!ht]
\includegraphics[width=19cm,angle=-90]{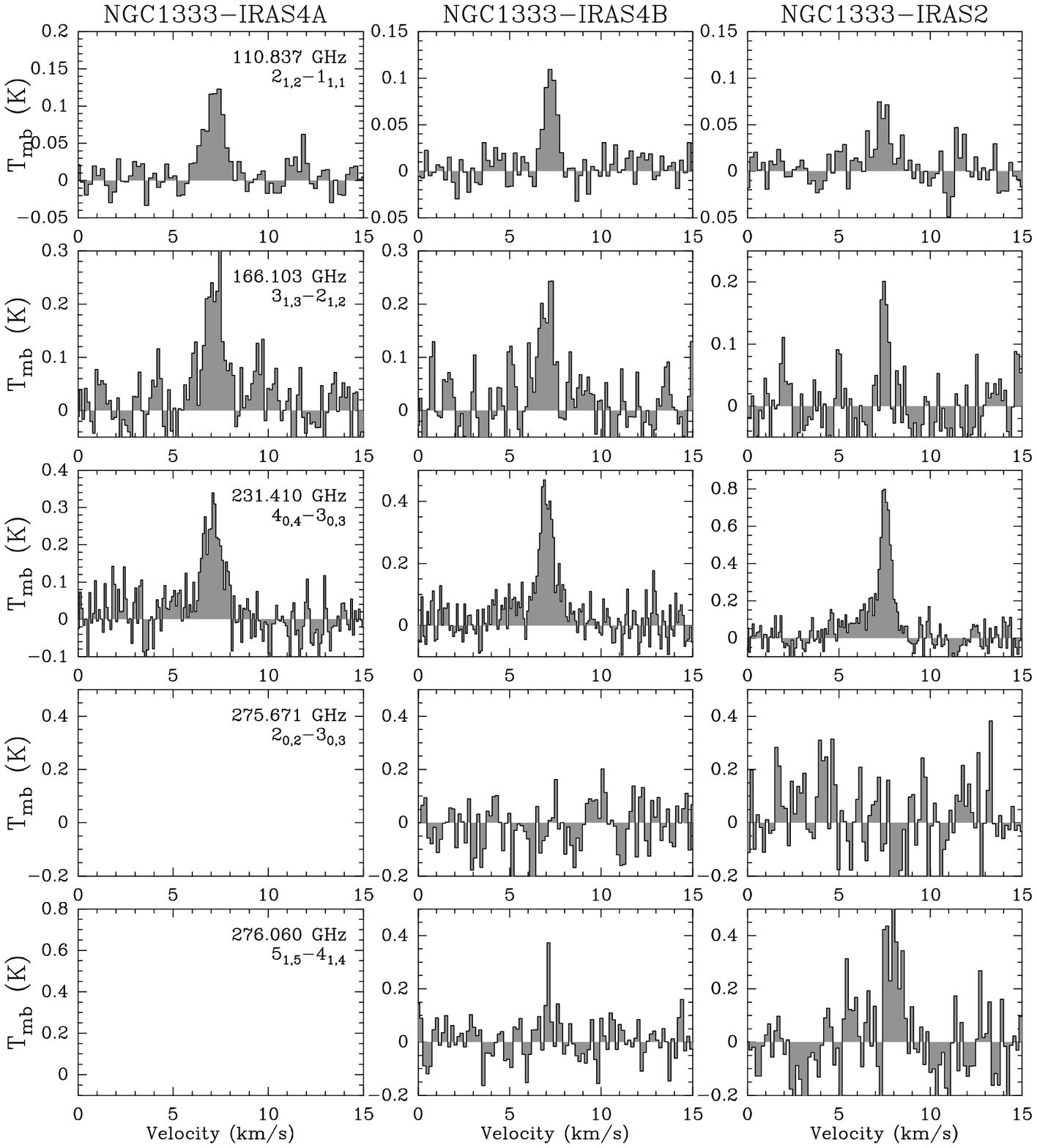}
\caption{D$_2$CO lines for IRAS4a, IRAS4b, IRAS2.}
\label{D$_2$COlines}
\end{figure*}

\begin{figure*}[!ht]
\includegraphics[width=19cm,angle=-90]{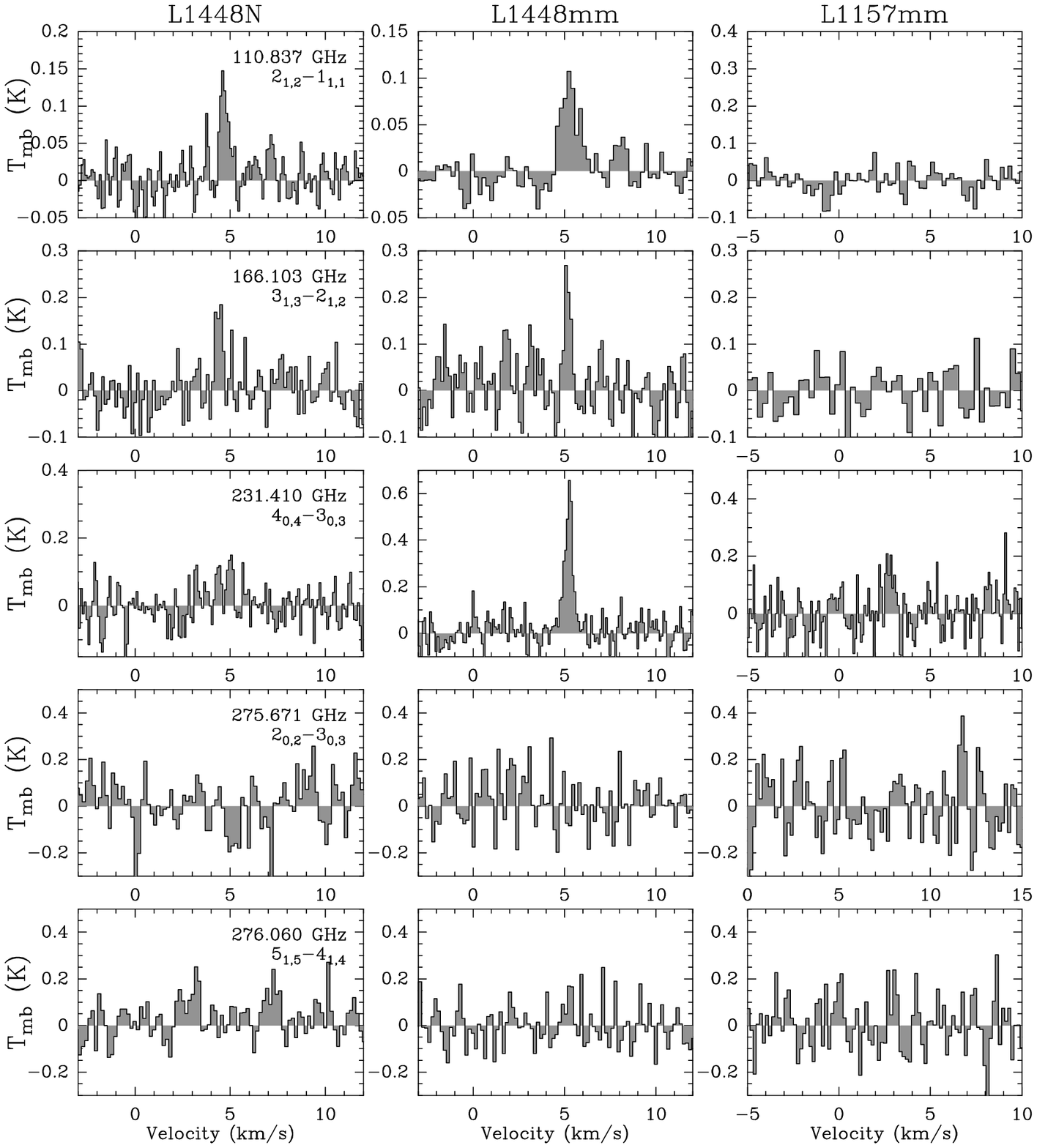}
\caption{D$_2$CO lines for L1448N, L1448mm, L1157mm}
\label{D$_2$COlines2}
\end{figure*}

\begin{figure*}[!ht]
\includegraphics[width=22cm,angle=-90]{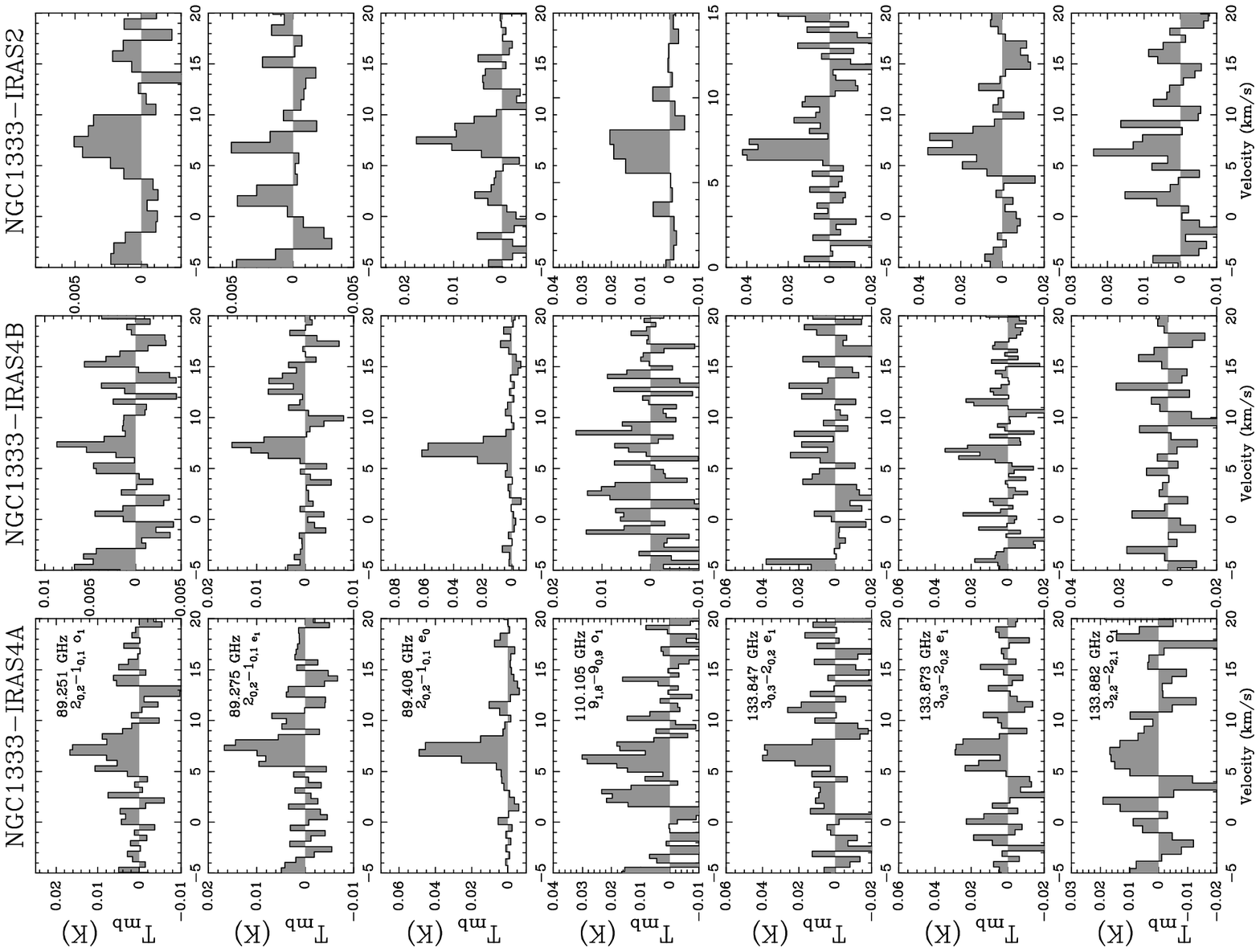}
\caption{CH$_2$DOH lines for IRAS4a, IRAS4b and IRAS2.}
\label{CH2DOHlines}
\end{figure*}

\begin{figure*}[!ht]
\includegraphics[width=22cm,angle=-90]{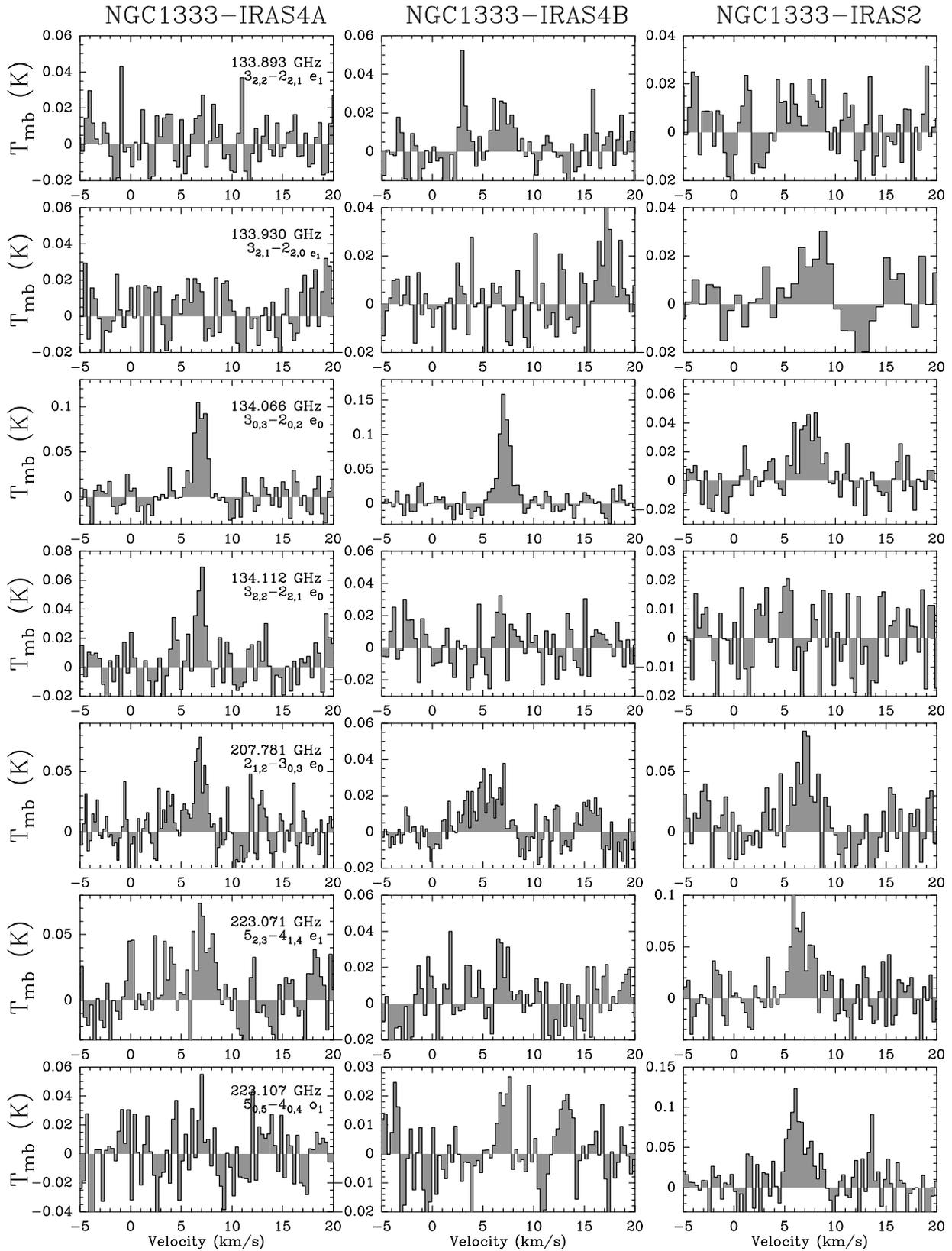}
\caption{CH$_2$DOH lines for IRAS4a, IRAS4b, IRAS2.}
\label{CH2DOHlines2}
\end{figure*}

Observed fluxes are listed in Tables \ref{flux}, \ref{flux1.5}, \ref{L1527} and \ref{flux2}.
Examples of observed spectra are shown in Fig. \ref{HDCOlines}, \ref{HDCOlines2}, \ref{D$_2$COlines}, \ref{D$_2$COlines2}, \ref{CH2DOHlines} and \ref{CH2DOHlines2}.

\subsection{Results}

We detected the two deuterated formaldehyde isotopes (HDCO and D$_2$CO) towards all 
the sources of our sample. Regarding methanol, only the three sources NGC1333$-$IRAS4A, 
$-$IRAS4B and $-$IRAS2 have good enough detections. This is consistent with the study by \citet{Maret05} 
that shows that these three sources are indeed the brightest ones for CH$_3$OH emission. CH$_2$DOH was detected in all sources in which it was searched for, but only the low-lying transition was detected in the case of L1448N, L1448mm and L1157mm. CH$_3$OD was detected only towards 
IRAS4A, IRAS4B, L1448mm and L1157mm, with only one transition detected in the two last sources.
Finally, doubly-deuterated methanol was detected towards IRAS4A, IRAS4B and IRAS2. Upper limits
were derived for the other sources. Despite the substantial integration time (420 minutes for L1448N, 
108 minutes for L1448mm and 264 minutes for L1157mm), upper limits on the fractionation are 
not very significant, mostly because of the low CH$_3$OH abundance in those sources \citep{Maret05}.  

The observed lines are relatively narrow, around 1 to 1.5 km\,s$^{-1}$ for deuterated formaldehyde and up to 3 km\,s$^{-1}$ for deuterated methanol. 

In particular for formaldehyde, the linewidths for the deuterated species are smaller than for 
the main isotopomer \citep{Maret04}. Indeed, the low energy H$_2$CO lines are probably contaminated 
by an outflow contribution, as discussed in \citet{Maret04}. The narrow lines emitted by the deuterated isotopomers of H$_2$CO suggest that the emission is dominated in this case by the cold outer envelope of the protostar, as expected from such lines with relatively low upper energies. This is in agreement with the observations of extended emission of D$_2$CO in the low-mass protostar IRAS16293$-$2422 \citep{Ceccarelli01}. This observation 
was interpreted as an indication that D$_2$CO comes from the evaporation of CO-rich ices that 
evaporate around 20\,K, {\it i.e.} at a lower temperature than polar ices.

For  methanol, the signal to noise ratio is too low to draw any firm conclusion,
but the deuterated lines also seem to be narrower than the main isotopomer, suggesting that 
the emission is also in this case dominated by the envelope contribution. However, the larger linewidths 
compared to formaldehyde as well as the higher rotational temperature (see next paragraph) 
suggests that it probes more deeply the envelope.

\section{Analysis}

The analysis based on an accurate model of infalling envelope of H$_2$CO and CH$_3$OH transitions towards the sample of low-mass protostars 
showed that the abundance of those two species jumps in the inner warm part of 
the envelope \citep{Maret04,Maret05}. This abundance jump has been attributed to the 
evaporation of H$_2$CO and CH$_3$OH from polar ices, in the region where the temperature is higher than 100\,K. In principle, a multifrequency analysis could be done for the CH$_2$DOH molecule, for
several lines have been observed in this molecule. However, the analysis could only be
done in the LTE approximation, for the collisional coefficients are not known.
Besides, since we cannot perform the analysis for the other molecules, we would not be able
to make any comparison anyway. We therefore decided not to implement this analysis
in this article. We will, on the contrary, use the rotational diagrams  technique, which gives the column density averaged on the source extent for optically thin and LTE lines. Both conditions are likely correct in our case, first because we do not expect particularly large column densities, second because the critical densities\footnote{For example, for CH$_2$DOH, the Einstein coefficients for the observed transitions lie in the 10$^{-6}$ to 10$^{-5}$ s$^{-1}$ range.  The collisional deexcitation rates 
are expected to be of the order of 10$^{-11}$\,cm$^3$\,s$^{-1}$, leading to critical densities 
of 10$^5$ to 10$^6$ cm$^3$. } for these lines are around 10$^5$ to 10$^6$ cm$^{-3}$  which are about the densities around these sources \citep{Maret04}.

The lines have been observed at several different frequencies, hence with different spatial resolution. Indeed, the beam size of the IRAM 30\,m telescope is 30$''$ at 83\,GHz and 9$''$ at 276\,GHz. If the source is smaller than the beam size of the observation, the derived column density must be corrected for the beam dilution. Unfortunately, we do not know for sure the size of the deuterated formaldehyde and methanol emission in the targetted protostars. In order to check whether and how much the uncertainty on the source size affects the derived column densities and fractionations, we discuss in the following paragraph a detailed study of the CH$_2$DOH data on IRAS4A.

\subsection{Can we constrain the size of the CH$_2$DOH emission ?}

We plotted the rotational diagrams of CH$_2$DOH towards IRAS4A (the source with the largest number of detected lines) assuming different 
sizes for the source emission: 10$''$, 15$''$, 20$''$, 25$''$ and 30$''$. The data have been acquired
at frequencies spanning 89\,GHz to 223\,GHz, implying beam sizes between 
30$''$ and 11$''$ (Table \ref{flux}). The different hypothesis on the source size 
will then correct differentially for the dilution depending on the frequency of the transition: 
if the beam size is smaller than the source, no correction is done, whereas dilution has to 
be corrected for observations with beam sizes bigger than the source. Methanol has the specificity to show no monotonic relation between the upper energy and the frequency of the transition (contrarily to CO for example), and thus the dilution correction has quite an unpredictable effect on the rotational diagram.  

For each assumed source size we calculated the reduced $\chi^2$ of the linear fit in the rotational diagram. If the assumed source size is the only origin of the scattering in the diagram, this reduced 
$\chi^2$ is minimum when the assumed source size is close to the real source size. 

The results of this study are presented in Table \ref{bid2}. The reduced $\chi^2$ 
decreases when increasing the source size, suggesting that the source is extended. Nevertheless, this whole study lies on three assumptions: the emission is homogeneous on the extent of the source, the source size is the same for all transitions, and scattering in the rotational diagram is only caused by the  dilution effect. These three hypothesis may not be valid. Indeed, the different transitions may be emitted by different regions, for instance the high energy transitions may originate in warmer and less extended regions than low lying transitions.
The scattering in the diagram may also be caused by opacity or non-LTE effects. It may thus be unrealistic to derive the source size by this method. Only interferometric observations may help to solve this issue.  

\begin{table*}[!htbp]
\begin{center}
\caption{Column densities for methanol isotopomers towards IRAS4A, derived by fixing the rotational temperature to the one derived for CH$_2$DOH, and ratios  between isotopomers.}
\label{bid2}
\begin{tabular}{c|ccc|cccccc}
\noalign{\smallskip}
\hline
\hline
\noalign{\smallskip}
Source  &  T$_{\rm rot}$ & CH$_2$DOH  &  $\chi^2_{\rm red}$ & CH$_3$OD & CHD$_2$OH  & \underline{CH$_3$OD} & \underline{CHD$_2$OH} \\
size  & (K) &  \footnotesize ($\times$10$^{14}$\,cm$^{-2}$) & & \footnotesize ($\times$10$^{13}$\,cm$^{-2}$) & \footnotesize ($\times$10$^{13}$\,cm$^{-2}$)  & CH$_2$DOH & CH$_2$DOH  \\
\noalign{\smallskip}
\hline
\noalign{\smallskip}
10$''$  & 27.1$\pm$1.3 & 4.3$\pm$0.4 & 13.5 & 3.1$\pm$0.7 & 11$\pm$1.7 & (7.2$\pm$1.8)$\times$10$^{-2}$ & 0.26$\pm$0.05 \\
15$''$  & 31.3$\pm$1.7 & 2.4$\pm$0.3 & 8.5 & 1.6$\pm$0.4 & 6.3$\pm$1.0  & (6.7$\pm$1.9)$\times$10$^{-2}$ & 0.26$\pm$0.05 \\
20$''$  & 36.2$\pm$2.3 & 1.8$\pm$0.2 & 5.2 & 1.2$\pm$0.3 & 5.0$\pm$0.9  & (6.7$\pm$1.8)$\times$10$^{-2}$ & 0.28$\pm$0.06 \\
25$''$  & 42.2$\pm$3.2 & 1.7$\pm$0.2 & 3.2 & 1.4$\pm$0.3 & 4.4$\pm$0.8  & (8.2$\pm$2.0)$\times$10$^{-2}$ & 0.26$\pm$0.06 \\
30$''$  & 46.2$\pm$3.8 & 1.7$\pm$0.2 & 2.9 & 1.5$\pm$0.4 & 3.9$\pm$0.7  & (8.8$\pm$2.6)$\times$10$^{-2}$ & 0.23$\pm$0.05 \\
\noalign{\smallskip}
\hline
\end{tabular}
\end{center}
\end{table*}

Nevertheless, it is worth studying the uncertainties led by the source size assumption on the column density of the various isotopomers. Table \ref{bid2} presents the column densities of methanol isotopomers versus the source size, obtained as follows. For CH$_3$OH, column densities can only be estimated accurately for sizes of emission smaller than 15$''$, because no observation was made with a beam larger than 15$''$. 
CH$_3$OD and CHD$_2$OH column densities were determined using the rotational temperature
derived for CH$_2$DOH, for which we have the largest number of transitions. Column densities change by a factor 2 to 3 according to the size of the source. On the contrary, the ratios between isotopomers vary only slightly with the source size ({\it cf.} Table \ref{bid2}). In the following, we will thus present the column densities assuming a source size of 10$''$. Finally, care should be taken in the use of the fractionation ratios relative to the main isotopomers (CH$_3$OH or H$_2$CO), as these species might be optically thick. On the contrary, ratios between~deuterated isotopomers are likely not to be affected by such problem and are thus reliable.

\subsection{Formaldehyde and methanol fractionation}

\begin{table*}[!htbp]
\begin{center}
\caption{H$_2$CO, HDCO and D$_2$CO rotational temperatures and column densities, assuming a 10$''$ source.}
\footnotesize
\begin{tabular}{cccccccc}
\noalign{\smallskip}
\hline
\hline
\noalign{\smallskip}
Source   &  \multicolumn{2}{c}{H$_2$CO$^a$}  &  \multicolumn{3}{c}{HDCO} &  \multicolumn{2}{c}{D$_2$CO}  \\
     &  T$_{\rm rot}$ & N$_{\rm tot}$ & T$_{\rm rot}$ &  N$_{\rm tot}$  & N$_{\rm tot}^b$  & T$_{\rm rot}$ & N$_{\rm tot}$  \\
 &  \scriptsize (K) & \scriptsize ($\times$10$^{14}$\,cm$^{-2}$) & \scriptsize (K) & \scriptsize ($\times$10$^{13}$\,cm$^{-2}$)   & \scriptsize ($\times$10$^{13}$\,cm$^{-2}$)  & \scriptsize (K) & \scriptsize ($\times$10$^{12}$\,cm$^{-2}$) \\
\noalign{\smallskip}
\hline
\noalign{\smallskip}
IRAS4A  & 20$\pm$2 & 1.7$\pm$0.4   & 7.3$\pm$0.3 & 3.2$\pm$0.6  & {\it 1.2$\pm$0.2}  & 5.0$\pm$0.3  & 19$\pm$4      \\
IRAS4B  & 38$\pm$6 & 1.7$\pm$0.5   & 8.2$\pm$0.5 & 1.9$\pm$0.4  & {\it 1.2$\pm$0.3}  & 7.9$\pm$0.5  & 7.0$\pm$1.5    \\
IRAS2   & 20$\pm$2 & 0.84$\pm$0.22 & 9.3$\pm$0.6& 1.3$\pm$0.3    & {\it 0.82$\pm$0.13} & 19.6$\pm$3.6 & 4.0$\pm$1.5   \\
L1448N  & 14$\pm$1 & 1.1$\pm$0.2 & 7.9$\pm$0.5 & 1.0$\pm$0.3  & {\it 0.46$\pm$0.057} & 4.9$\pm$0.8  & 8.2$\pm$4.4    \\
L1448mm & 16$\pm$1 & 0.42$\pm$0.11 & 8.7$\pm$0.6 & 1.1$\pm$0.3  & {\it 0.57$\pm$0.063}  & 5.7$\pm$0.3  & 9.1$\pm$2.3    \\
L1157mm & 12$\pm$1 & 0.20$\pm$0.06 & 10.0$\pm$0.9 & 0.29$\pm$0.09  & {\it 0.24$\pm$0.034}& -            & $\le$ 1.5      \\
L1527   & 12$\pm$1 & 0.42$\pm$0.17 & 4.8$\pm$0.3 & 6.0$\pm$1.8 &{\it 1.1$\pm$0.16} & 5.1$\pm$0.3  & 14.9$\pm$4.2       \\ 
\noalign{\smallskip}
\hline
\noalign{\smallskip}
\end{tabular}
\label{tableform}
\end{center}
$^a$\,Reanalysis of the data from \citet{Maret04} with the rotational diagram method ({\it cf.} text).
$^b$\,Assuming the same rotational temperature as H$_2$CO.
\end{table*}

\begin{figure}[!htbp]
\begin{center}
\includegraphics[width=0.5\textwidth]{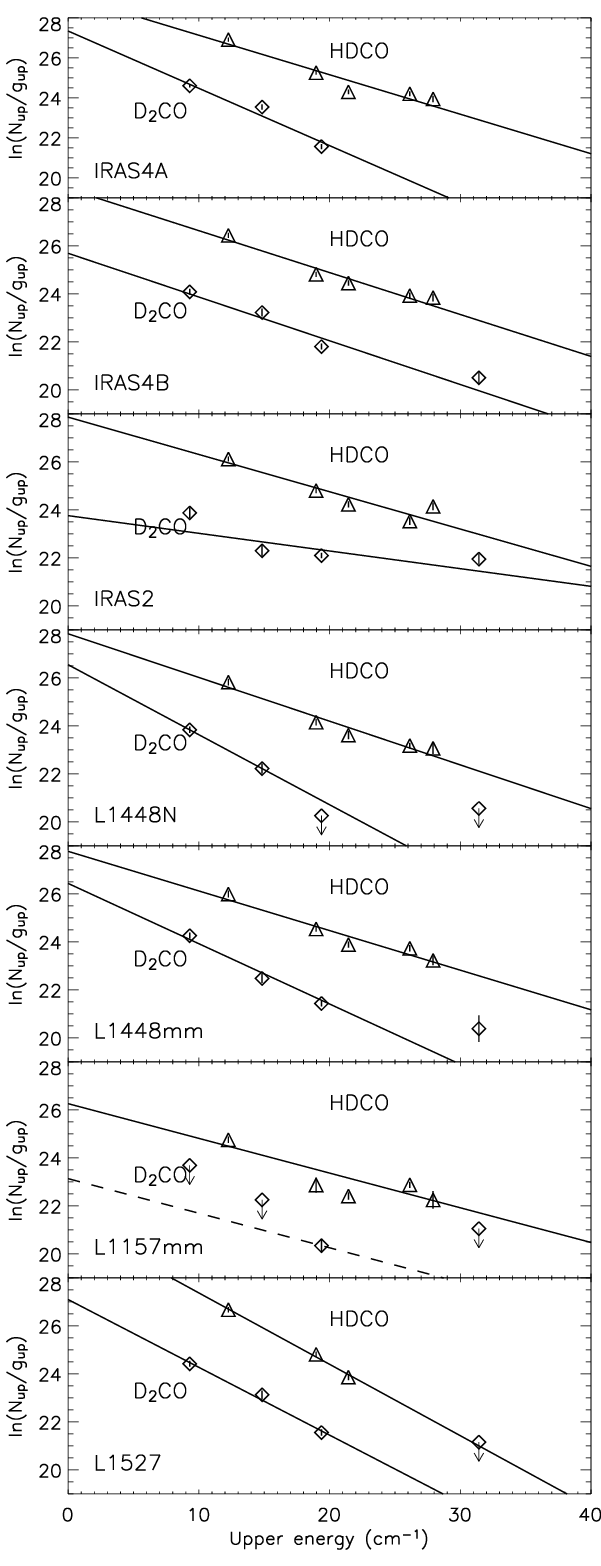}
\end{center}
\caption{Rotational diagrams for deuterated formaldehyde for the seven studied sources, assuming 10$''$ for the source size. Error bars correspond to the error bars on the flux as stated in Tab. \ref{flux}, \ref{flux1.5} and \ref{L1527}.}
\label{form_all}
\end{figure}

\begin{table*}[!htbp]
\begin{center}
\caption{Methanol rotational temperatures and column densities, assuming a 10$''$ source size.}
\begin{tabular}{ccccccc}
\noalign{\smallskip}
\hline
\hline
\noalign{\smallskip}
Source & \multicolumn{2}{c}{CH$_3$OH}  & \multicolumn{2}{c}{CH$_2$DOH} & CH$_3$OD & CHD$_2$OH \\
     & T$_{\rm rot}$ &  N$_{\rm tot}$ & T$_{\rm rot}$ &  N$_{\rm tot}$ &  N$_{\rm tot}$ & N$_{\rm tot}$ \\
 & \footnotesize (K) & \footnotesize ($\times$10$^{14}$\,cm$^{-2}$) & \footnotesize (K) & \footnotesize ($\times$10$^{14}$\,cm$^{-2}$) & \footnotesize ($\times$10$^{13}$\,cm$^{-2}$)  & \footnotesize ($\times$10$^{14}$\,cm$^{-2}$) \\
\noalign{\smallskip}
\hline
\noalign{\smallskip}
IRAS 4A & 38.0$\pm$3.2  & 6.9$\pm$1.4  & 27.1$\pm$1.3     & 4.3$\pm$0.4 & 3.1$\pm$0.7$^c$ & 1.1$\pm$0.17$^c$ \\
IRAS 4B & 84.9$\pm$16.8 & 8.0$\pm$2.7  & 15.6$\pm$0.5     & 2.9$\pm$0.2 & 1.1$\pm$0.20$^c$ & 0.90$\pm$0.10$^c$ \\
IRAS 2  & 207$\pm$48.1  & 10.1$\pm$3.7     & 55.3$\pm$4.9     & 5.2$\pm$0.8 & $\le$ 8.0 & 2.1$\pm$0.4$^c$ \\
L1448N  & 27.5$\pm$4.5  & 1.2$\pm$0.47 & 27.5$\pm$4.5$^a$ & 2.1$\pm$0.6$^b$ & $\le$ 8.0 & $\le$ 10 \\
L1448mm & 76.3$\pm$19.1 & 1.6$\pm$0.70 & 76.3$\pm$19.1$^a$& 11.3$\pm$4.7$^b$ & 40$\pm$16$^b$ & $\le$ 6.3 \\
L1157mm & 112$\pm$201   & 1.9$\pm$5.2  & 112$\pm$201$^a$  & 10.3$\pm$27.7$^b$ & 22$\pm$66$^b$ & $\le$ 9.6 \\
\noalign{\smallskip}
\hline
\noalign{\smallskip}
\end{tabular}
\label{source10}
\end{center}
$^a$ Temperature fixed to the rotational temperature of CH$_3$OH.\\
$^b$ Should be taken as an upper limit, as the rotational temperature is likely to be overestimated.\\
$^c$ Temperature fixed to the rotational temperature of CH$_2$DOH.\\
\end{table*}

Figure \ref{form_all} presents the rotational diagrams derived for the deuterated isotopomers of formaldehyde towards the seven sources of the sample assuming 10$''$ source sizes. Table \ref{tableform} lists the derived rotational temperatures and column densities. For consistency with the analysis of HDCO and D$_2$CO, the H$_2$CO column density has been recomputed with the rotational diagram method from the data of \citet{Maret04}, with the ortho/para ratio fixed to its statistical value (contrarily to the study of \citet{Maret04} where this ratio was considered as a free parameter), and taking into account the line opacities 
for the transitions where H$_2$$^{13}$CO has been observed. Note that the column densities inferred by treating the ortho/para ratio as a free parameter are less than a factor 2 different from our values \citep{Maret04}.
The derived rotational temperatures for HDCO and D$_2$CO 
are in some cases different from the one derived for H$_2$CO, but it should be noticed that the transitions only span a small interval of upper energies and thus do not constrain very well the rotational temperature. In order to get a sense of the uncertainty implied by these different rotational temperatures, we also computed the HDCO column densities assuming the same rotational temperature as H$_2$CO (cf Table \ref{tableform}). Results were found to be in most cases within a factor of two uncertainties, and the fractionation ratios mostly appear to be consistent within the error bars (cf Table \ref{recap}, footnote a).

\begin{figure}[!htbp]
\begin{center}
\includegraphics[width=0.5\textwidth]{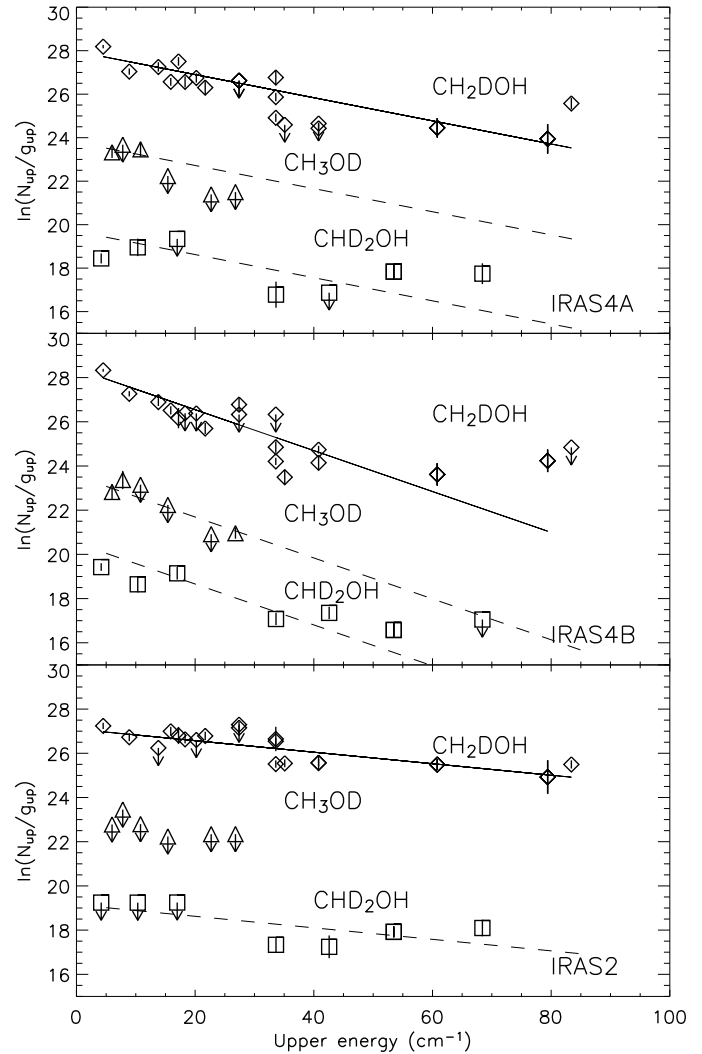} 
\end{center}
\caption{Rotational diagrams for deuterated methanol in IRAS4A, IRAS4B and IRAS2 (for the sake of clarity the curves for CH$_3$OD  and  CHD$_2$OH have been translated by -2 and -7 along the y-axis). Error bars correspond to the error bars on the flux as stated in Tab. \ref{flux} and \ref{flux1.5}. Solid lines (CH$_2$DOH) correspond to fits to the data, with two free parameters (T$_{\rm rot}$ and N$_{\rm tot}$). Dashed lines correspond to fits with only N$_{\rm tot}$ as a free parameter, and T$_{\rm rot}$ set to the value derived for CH$_2$DOH (cf text).} 
\label{rotdiag_3en1_all}
\end{figure}

Fig. \ref{rotdiag_3en1_all} presents the rotational diagrams of the deuterated forms of methanol 
for the three sources IRAS4a, IRAS4b and IRAS2 respectively, again assuming a source size of 10$''$. The CH$_3$OD and CHD$_2$OH column densities were computed assuming the same rotational temperature as 
CH$_2$DOH, for which many more lines were observed in all sources.
As for formaldehyde, the CH$_3$OH column density was recomputed from \citet{Maret05} 
using rotational diagrams. 

\bigskip 
\bigskip
\bigskip


\section{Discussion}

In this section, we discuss the results of the observation of deuterated formaldehyde and methanol towards the sample of low-mass class 0 protostars. 
Table \ref{recap} summarizes the observed deuterium fractionations. For comparison, we added the fractionations measured in IRAS16293 \citep{Loinard01, Parise02,Parise04}. \citet{Roberts02a} 
observed the HDCO fractionation in a number of protostellar cores, including L1448mm and L1527, 
using the Kitt Peak 12\,m telescope. They found values sensibly lower than ours in these two sources (resp. 0.069 and 0.066). This may come from the fact that the 12\,m telescope beam size is 2.5 times bigger than the 30\,m, so that their observations encompass the surrounding cloud where the fractionation is expected to be lower. 

\begin{table*}[!htbp]
\footnotesize
\begin{center}
\caption{Deuterium fractionation for formaldehyde and methanol, for an assumed source size
of 10$''$. Error bars are associated with a 90\% confidence interval (see appendix).}
\label{recap}
\begin{tabular}{lccccccc}
\noalign{\smallskip}
\hline
\hline
\noalign{\smallskip}
Source & ~\underline{HDCO}$^a$ & \underline{D$_2$CO} & \underline{D$_2$CO} & \underline{CH$_2$DOH} & \underline{CH$_3$OD} & \underline{CHD$_2$OH}& \underline{CHD$_2$OH}\\
& H$_2$CO & H$_2$CO & HDCO & CH$_3$OH & CH$_3$OH & CH$_3$OH & CH$_2$DOH \\
\noalign{\smallskip}
\hline
\noalign{\smallskip}
IRAS4A & 0.20$_{-0.09}^{+0.13}$ & 0.12$_{-0.05}^{+0.08}$ & 0.62$_{-0.26}^{+0.33}$ & 0.65$_{-0.21}^{+0.30}$ & 0.047$_{-0.021}^{+0.029}$ & 0.17$_{-0.06}^{+0.09}$ & 0.26$_{-0.07}^{+0.08}$ \\
\noalign{\smallskip}
IRAS4B & 0.13$_{-0.06}^{+0.10}$ & 0.046$_{-0.024}^{+0.038}$ & 0.39$_{-0.17}^{+0.23}$ & 0.43$_{-0.20}^{+0.38}$ & 0.016$_{-0.008}^{+0.016}$ & 0.13$_{-0.07}^{+0.12}$ & 0.31$_{-0.06}^{+0.07}$ \\
\noalign{\smallskip}
IRAS2  & 0.17$_{-0.08}^{+0.12}$ & 0.052$_{-0.034}^{+0.048}$ & 0.33$_{-0.21}^{+0.28}$ & 0.62$_{-0.33}^{+0.71}$ & $\le$ 0.08 & 0.25$_{-0.14}^{+0.29}$ & 0.41$_{-0.15}^{+0.19}$ \\
\noalign{\smallskip}
L1448N & 0.094$_{-0.050}^{+0.062}$ & 0.077$_{-0.069}^{+0.079}$ & 0.93$_{-0.83}^{+1.14}$ & $\le$ 1.8 & $\le$ 0.67 & $\le$ 8.3 & -- \\
\noalign{\smallskip}
L1448mm& 0.29$_{-0.15}^{+0.23}$ & 0.24$_{-0.12}^{+0.18}$ & 0.91$_{-0.48}^{+0.73}$ & $\le$ 7.1 & $\le$ 2.5 & $\le$ 3.9 & -- \\
\noalign{\smallskip}
L1157mm& 0.16$_{-0.10}^{+0.15}$ & $\le$ 0.08    & $\le$ 0.52 & $\le$ 5.4 & $\le$ 1.2 & $\le$ 5.1 & -- \\
\noalign{\smallskip}
L1527  & 1.7$_{-1.1}^{+2.6}$ & 0.44$_{-0.29}^{+0.60}$ & 0.28$_{-0.16}^{+0.25}$ & -- & -- & -- & -- \\
\noalign{\smallskip}
\hline
\noalign{\smallskip}
IRAS16293 & 0.15$\pm$0.07 & 0.05$\pm$0.025 & 0.3$\pm$0.2  &$^b$0.37$_{-0.19}^{+0.38}$ &  $^b$0.018$_{-0.012}^{+0.022}$ & $^b$0.074$_{-0.044}^{+0.084}$ & $^b$0.21$_{-0.10}^{+0.11}$ \\
\noalign{\smallskip}
\hline
\noalign{\smallskip}
\end{tabular}
\end{center}
$^a$\,HDCO/H$_2$CO ratios obtained when assuming the same rotational temperature for HDCO as for H$_2$CO are resp. 0.075$_{-0.031}^{+0.047}$, 0.080$_{-0.044}^{+0.068}$, 0.11$_{-0.04}^{+0.07}$, 0.043$_{-0.015}^{+0.021}$, 0.15$_{-0.06}^{+0.10}$, 0.14$_{-0.06}^{+0.11}$, 0.33$_{-0.18}^{+0.42}$ for the 7 sources.\\
HDCO/H$_2$CO ratios obtained assuming a 15$''$ source are resp. 0.14$_{-0.07}^{+0.11}$, 0.07$_{-0.05}^{+0.09}$, 0.14$_{-0.07}^{+0.11}$, 0.08$_{-0.04}^{+0.06}$, 0.23$_{-0.14}^{+0.19}$, 0.19$_{-0.12}^{+0.22}$, 0.89$_{-0.63}^{+1.43}$.
They are thus consistent with the ratios derived for a 10$''$ source, within the error bars.

$^b$\,Error bars recomputed from \citet{Parise04} for a 90\% confidence interval (see appendix).
\end{table*}

The main result of this observational study is the discovery of a high 
fractionation ratio for formaldehyde and methanol for each source where 
these molecules were detected. The measured fractionation ratios are indeed 
similar to those observed in IRAS 16293. In the sources where the 
isotopes were not detected, the derived upper limits do not exclude
a similar deuteration. The present observations confirm that IRAS16293 is not an 
exception regarding deuteration. Other studies also confirm that its chemistry is not peculiar.
Indeed \citet{Bottinelli04a} observed complex molecules in the Hot Core of IRAS4A, molecules that 
were detected for the first time in a low-mass protostar towards IRAS16293 \citep{Cazaux03}.

We explore in the following paragraph the correlations between the deuteration and other parameters. We then compare the observed fractionations to the fractionations predicted by grain chemistry models.

\subsection{Correlations}

\begin{figure*}[!htbp]
\centering
\includegraphics[width=0.8\textwidth]{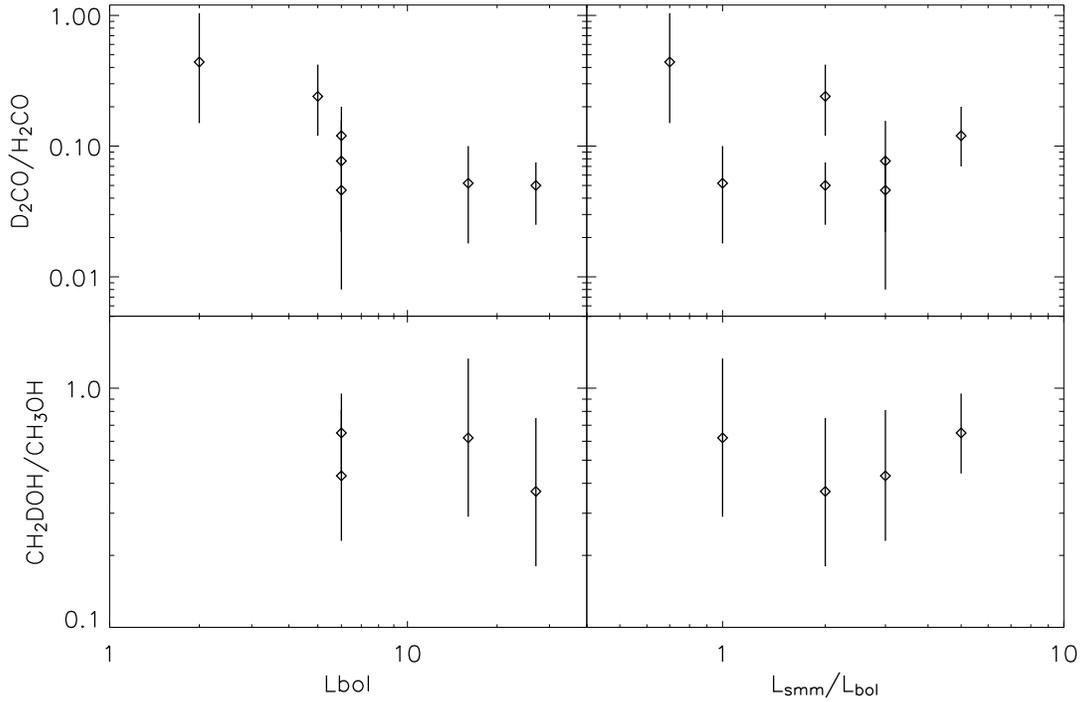}   
\caption{Fractionation of formaldehyde and methanol as a function of bolometric luminosity and the L$_{smm}$/L$_{bol}$ \citep[from][]{Andre00}. }
\label{correl_bid}
\end{figure*}

Figure \ref{correl_bid} shows the measured D$_2$CO/H$_2$CO and CH$_2$DOH/CH$_3$OH as
function of the bolometric luminosity and the L$_{smm}$/L$_{bol}$ \citep[from][]{Andre00}. The figures show no obvious correlations between the measured fractionation and these two quantities, despite a variation of one order of magnitude in the D$_2$CO/H$_2$CO ratio. This is indeed consistent with the hypothesis that the deuteration is a memory of precollapse phase, rather than a present day product \citep[e.g.][]{Ceccarelli01}. If one cannot therefore use the fractionation to distinguish the evolutionary status of the sources, it is likely that the fractionation gives insights of the precollapse phase of each source. 

\begin{figure}[!htbp]
\centering
\includegraphics[width=0.5\textwidth]{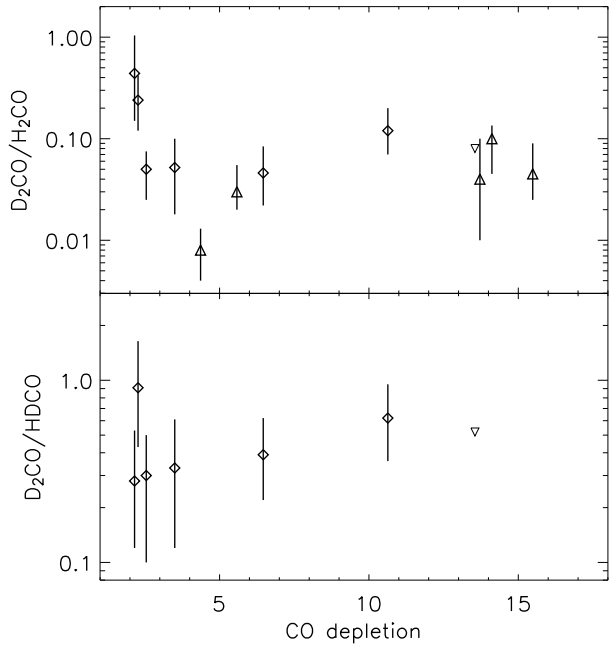}
\caption{Doubly-deuteration of formaldehyde versus CO depletion. Class 0 sources (this work) appear as diamonds and prestellar cores \citep{Bacmann02} as triangles. The reversed triangle is the upper limit for L1157mm (this work).}
\label{correl}
\end{figure}

Figure \ref{correl} presents the D$_2$CO/H$_2$CO fractionation ratio (as well as the D$_2$CO/HDCO ratio in order to get rid of any potential opacity effect) versus 
the CO depletion in the sources of our sample. The CO depletion was computed 
from the CO densities in the outer envelope derived by \citet{Jorgensen02}, and
assuming a canonical CO abundance of 8.4$\times$10$^{-5}$ \citep{Frerking82}. 
We also plotted in this figure the result for prestellar cores observed by \citet{Bacmann02}. 

The first thing to notice is that Class 0 sources and prestellar cores with similar CO depletion factors do present also similar D$_2$CO/H$_2$CO ratios. This is a further suggestion that the deuteration of formaldehyde in Class 0 sources occurred during the precollapse phase, and/or that the physical conditions in the outer envelopes of Class 0 sources are very similar to prestellar cores. 
On the other hand, and surprising enough, the two sources with the largest D$_2$CO/H$_2$CO ratio are those with the lowest CO depletion factor. Indeed chemical models \citep[e.g. ][]{Roberts00b} 
predict that formaldehyde fractionation should increase with CO depletion.
However, we inquire in the following if this could be caused by the optical thickness of H$_2$CO, leading to artificially high D$_2$CO/H$_2$CO ratios. Both sources (L1527 and L1448mm) present evidences of very optically thick CO emission.
A study of C$^{18}$O and C$^{17}$O emission \citep{Jorgensen02} towards a sample including all our sources showed that L1527 and L1448mm are the sources for which C$^{18}$O is the most optically thick \citep{Jorgensen02}.
H$_2$CO is thus more likely to be optically thick \citep[as suggested by the correlation between
the H$_2$CO and CO abundances, ][]{Maret04}.
\citet{Maret04} have observed some transitions from H$_2^{13}$CO in a few sources. Unfortunately, 
no line was observed for L1527, and only an upper limit was derived for one line on L1448mm 
($\tau$ $<$ 2). In the case of L1448mm, we thus might have underestimated the H$_2$CO column density by a factor as high as $\frac{\tau}{1-e^{\tau}}$ $\sim$ 2.3. For L1527, the opacity is likely to be even larger as suggested by the C$^{18}$O/C$^{17}$O measured by \citet{Jorgensen02}.
However, we conclude from this study that this opacity effect cannot fully account for the discrepancy of 2 orders of magnitude observed in D$_2$CO/H$_2$CO compared to prestellar cores. Indeed, the D$_2$CO/HDCO, which is supposed to be less affected by opacity effects than the ratio with the main isotopomer, is still too high for one of the sources (L1448mm). 
Hence, the conclusion is while there seems to be a correlation between CO depletion and fractionation in prestellar cores, no such correlation is apparent in our sample of YSO (no correlation is found either  between the fractionation and the N$_2$H$^+$ abundance, tracing the N$_2$ depletion). This difference may imply that the CO depletion in the circumstellar envelopes is affected by outgassing due to the heating by the newly formed star.

Finally, our analysis, based on rotational diagrams, allows only to estimate the fractionation averaged on the beam. It will be important to understand what the contribution of the different regions (warm envelope, cold envelope, etc) is. New insights in this respect may only be possible by interferometric observations, to observe the relative fractionation of formaldehyde and methanol in the two regions : hot core and outer envelope.

\subsection{Comparison to grain chemistry models}

Grain surface chemistry is thought to be responsible for the formation of the abundant methanol 
observed in hot cores, for two main reasons. First, gas-phase chemistry models fail to reproduce 
the CH$_3$OH abundance by 2 to 3 orders of magnitude in Orion \citep{Menten88}. Recent investigations have confirmed that no gas-phase process can efficiently form methanol \citep{Herbst05}. Second,
methanol is one of the most abundant species observed in the grain mantles (after water), with 
abundances as high as 30\% of the water abundance \citep{Dartois99}. 

Methanol is thought to be formed on the grains by successive hydrogenations of CO. An intermediate product of these reactions is formaldehyde. Deuterium fractionation 
studies of formaldehyde and methanol thus provide a useful tool for confirming the grain chemistry scenario. 
To understand if formaldehyde and methanol are formed simultanuously on the grains,
and to unveil a possible contribution of gas-phase processes in the formation of formaldehyde,
we compared the observed fractionations to the predictions of the grain chemistry model
from \citet{Stantcheva03}. This model is based on the resolution of the rate equations governing the abundance of each species on the grain, and uses the master equation approach in the cases 
where only a small number of molecules are present on the surface.

\begin{figure*}[!htbp]
\includegraphics[width=\textwidth]{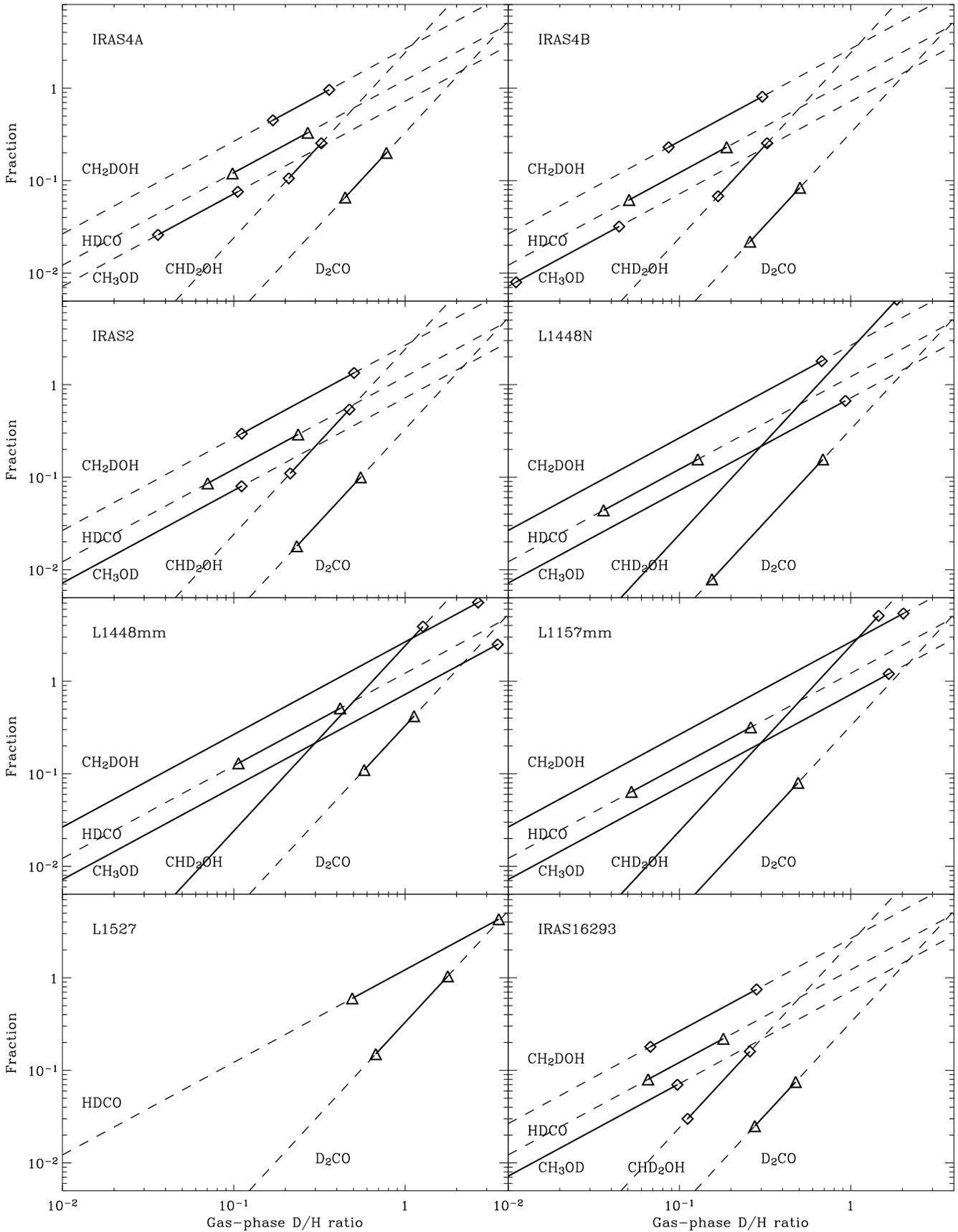}
\caption{Comparison between the observed fractionations (triangles are for formaldehedyde species whereas diamonds are for methanol species) towards the seven sources of our sample (solid lines) and the predictions of the grain model \citep[dashed lines,][]{Stantcheva03}. IRAS16293 has been added in the last panel for comparison.}
\label{grain_survey}
\end{figure*}

Fig.~\ref{grain_survey} presents, in dashed lines, the HDCO, D$_2$CO, CH$_2$DOH, CH$_3$OD et CHD$_2$OH fractionations predicted by the grain model, as a function of the gas-phase atomic 
D/H ratio at the time of mantle formation. The observed fractionations with their error bars have been superimposed for each source. This allows to infer the required D/H ratio required for the formation of each molecule.
 
For a comparison, Fig.~\ref{grain_survey} also presents the same study for IRAS16293 \citep{Ceccarelli98, Loinard00, Parise04}.

A comparison between formaldehyde on the one hand and methanol on the other hand
must be led with care as it is very dependent on the possible opacity of the main isotopomers 
H$_2$CO and CH$_3$OH. On the opposite, a comparison between different deuterated isotopomers of methanol should not suffer from such problem. We attribute 
to this opacity issue the fact that, depending on the source, either HDCO or D$_2$CO is compatible with methanol fractionation.

\bigskip

Several conclusions can be drawn from this comparison~:\\

$\bullet$ Whenever they are correctly constrained, the CH$_2$DOH and CHD$_2$OH fractionations are consistent, {\it i.e.} they have a common range of possible atomic D/H ratio (IRAS16293, IRAS4A, IRAS4B et IRAS2). 
That had already been observed for IRAS16293 \citep{Parise02,Parise04}, for which CD$_3$OH 
fractionation was also consistent with the latter two isotopomers.
The present new observations thus add further support to the hypothesis that methanol is formed on the grain surfaces. \\

$\bullet$ Whenever it is contrained, CH$_3$OD appears to be underabundant with respect to the other methanol isotopomers. This had already been noticed for the case of IRAS16293 by \citet{Parise02},  who  concluded that CH$_3$OD may be destroyed 
in the gas phase by protonation more rapidly than the other isotopomers, owing to the strong electronegativity of the oxygen atom. This conclusion was also suggested by 
the theoretical study of \citet{Osamura04}. However, some very recent laboratory experiments suggest 
another possibility. Indeed, \citet{Nagaoka05} show that deuterated methanol species with an OD bond are very inefficiently formed on ice surfaces when exposing CH$_3$OH to H and D atoms, while isotopomers with CD bonds are abundantly formed. This might also explain the non-detection of HDO in ices of protostellar envelopes \citep{Dartois03,Parise03} and the fractionation of water observed to be one order of magnitude lower than CH$_2$DOH/CH$_3$OH in the hot corino of IRAS16293, in agreement with these solid state observations \citep{Parise05a}.\\

$\bullet$ HDCO appears to be less abundant than D$_2$CO with respect to the model predictions (except for L1157mm, but for this source, the error bars are important). No preferential destruction in the gas phase could explain this discrepancy (as it was the case for CH$_3$OD). 
This might be caused by H-D substitution via hydrogen abstraction reactions (${\rm HDCO + H \rightarrow DCO +H_2}$), which are not included in the considered chemistry model. These reactions would favour the formation of D$_2$CO by abstracting an hydrogen to HDCO and deuterating DCO \citep{Tielens83}.\\
Alternatively, this might be a hint that formaldehyde formation and fractionation is not dominated by grain chemistry processes. We will discuss more thoroughly the gas-phase formation and fractionation of formaldehyde in the next section.

\subsection{A gas-phase formation of formaldehyde ?}

As discussed in the previous section, there is now good evidence that methanol is formed on the grains, and the data of this survey add further support to this hypothesis (as presented in the previous paragraph). Yet, the situation is not so clear for formaldehyde. After discussing the spatial origin of the HDCO and D$_2$CO emission, we examine the implications of the present observations on the formaldehyde formation process.

Contrarily to methanol that shows quite high rotational temperatures, and thus may be arising
in the hot corinos, deuterated formaldehyde shows low rotational temperatures, rather pointing to an origin in the cold envelope of the protostars. The study by \citet{Ceccarelli01} of the
distribution of the D$_2$CO  4$_{0,4}-$3$_{0,3}$ line (E$_{up}$=19.4\,cm$^{-1}$, a rather low energy line) towards IRAS16293-2422 shows that it is extended
over the entire envelope, and that the D$_2$CO/H$_2$CO ratio is also large across the
envelope. These authors also noticed that while the D$_2$CO emission is
well correlated with the continuum emission tracing the envelope, it does not show any
enhancement in correspondence of the shocks sites in the region caused by the interface
between the outflow and the envelope itself. They concluded that D$_2$CO is abundant
{\it in } the cold outer envelope, where the dust temperature is lower than 100\,K.
Of course, this does not exclude the possibility of abundant D$_2$CO also in the hot corino
region, but the available observations were unable to detect it. Note that Fuente et al. (2005)
found, from inferferometric observations, abundant D$_2$CO in the hot core of the intermediate mass protostar NGC7129-FIRS2.
Similarly to the case of IRAS16293-2422, the present observations are unable to disentangle
a possible contribution from the hot corino, as well as from the shocks around these sources.
But they cannot exclude these contributions either.
Given the uncertainty on the location of the D$_2$CO, in the following we analyze both
possible situations, hot corino versus cold envelope. 

The high observed deuterium fractionation of formaldehyde \citep[][ and this paper]{Ceccarelli98, Loinard01}, if arising in the hot corino, excludes an efficient production of formaldehyde in the warm gas, for which a much lower deuteration would be expected. Hence these high formaldehyde abundance and  fractionation must reflect the cold phase preceeding the hot core phase accompanying the formation of the protostar.
Gas phase models typically produce formaldehyde abundance which range from about 10$^{-6}$ at early times to 10$^{-8}$ at late times (e.g. in steady state). Observed abundance in cold dark clouds are typically 10$^{-8}$ and in good agreement with the latter value. However, abundances are observed to be higher in the hot corino around protostars \citep[$\sim$ 10$^{-7}$-10$^{-6}$,][]{Maret04, Maret05}. Hence, models based upon gas phase production of the formaldehyde in hot cores have to selectively deplete the formaldehyde in order to keep the abundance in the dark phase low. While all species are expected to freeze out equally efficiently at the low temperature of dust grains in dark clouds, perhaps, CO is returned to the gas phase by evaporation but formaldehyde is left behind. Then, in order to increase the formaldehyde abundance by a factor of 100, the CO has to cycle a hundred times between the gas phase and the ices. This seems a bit contrived.
The only gas phase possibility seems therefore formation at early times, when H$_2$CO abundance is 10$^{-6}$, and immediate depletion and cold storage in unreactive ices. This seems to be in contradiction with the depletion studies of dark cloud cores \citep{Tafalla02}.
Furthermore, the bulk of the formaldehyde would be formed under undepleted conditions and hence the D$_2$CO and HDCO fractionation is difficult to understand \citep{Roberts03}.

We now study the implications of our observations if the HDCO and D$_2$CO mainly originate in the cold envelope. In this case, 
we must inquire whether gas phase processes at present times can account for their formation. The discrepancy of our fractionation observations for formaldehyde with the grain chemistry models (namely the fact that HDCO seems to be steadily underabundant compared to D$_2$CO) might indeed reveal that the formation of formaldehyde is not dominated by H and D additions to CO on grain surfaces. In this cold gas, the absolute H$_2$CO abundance is well explained by gas-phase models. 
Models of formaldehyde fractionation in prestellar cores --whose chemical and physical conditions are similar to those in the outer cold envelopes of Class 0 sources \citep{Maret04,Jorgensen05}-- seem approximatively to reproduce the observed ratios \citep{Roberts04}. However, the predictions depend on some key reactions, whose rates, products, and branching ratios are relatively uncertain, and recent studies seem to show that gas-phase models may have more difficulties to reproduce the high deuteration than expected. Indeed, \citet{Osamura05} recently presented quantum chemical calculations of the protonation/deuteration of H$_2$CO and deuterated isotopologues. These computations show that the deuteration of H$_2$CO by H$_3^+$ and deuterated isotopomers will mainly happen on the oxygen end of the molecule, due to the higher electronegativity of oxygen. It is generally expected that dissociative electron recombination of H$_2$COD$^+$ will not lead to the formation of HDCO, so that the fractionation due to reactions of H$_2$CO with H$_2$D$^+$ and other isotopomers is quite inefficient. However, given the recent surprising experimental results on the product distribution in the
dissociative recombination reaction of N$_2$H$^+$ and CH$_3$OH$_2^+$ \citep{Geppert06}, an experimental verification of the product distribution of this
key reaction (H$_2$COD$^+$ + e$^-$) would be very valuable. Another remaining
possibility in this context is that H$_2$CO is formed in the gas-phase and then frozen out on dust grains where it can be later fractionated by abstraction reactions on the grain surfaces. Some efficient desorption processes are then required to release the molecules back to the gas-phase. \\

In summary, pure gas-phase models seem to face problems to explain both formaldehyde absolute abundance and fractionation at the same time. Unless the dissociative recombination of H$_2$COD$^+$ leads to the formation of HDCO, H-D substitutions on grains seem to be required to account for the high fractionation if formaldehyde is formed by gas phase processes. 

\section{Conclusion}

We have presented observations of deuterated formaldehyde and methanol 
towards a sample of low-mass protostars. We detected HDCO, D$_2$CO
and CH$_2$DOH in all sources. However, CH$_3$OD and CHD$_2$OH were only detected 
towards 2 and 3 sources respectively.
We analyzed the data using population diagrams. These observations show that 
IRAS16293$-$2422 is not an exception concerning deuteration. The fractionations
derived towards the source of our sample are indeed similar to those observed in IRAS16293.

These observations are useful to pin down the contributions of grain surface and gas phase chemistry for the formation of formadehyde and methanol. It appears that methanol fractionation is consistent with grain surface schemes, whereas formaldehyde formation cannot be fully constrained from these data. Two possibilities arise, either that formaldehyde formation is dominated by gas-phase reactions, later fractionated on the grains through abstraction reactions, or that formaldehyde is formed on the grains, with abstraction reactions altering the expected D$_2$CO/HDCO ratio.

This study is limited by different factors. All the transitions were observed with different sized beams,
with a best spatial resolution of 10$''$. It is in this case very difficult to derive the abundance of the 
molecules if one does not know the source size. Moreover, the spatial resolution is too poor to 
derive the fractionation in the different regions of the sources (warm inner envelope, cold outer envelope, outflows ...). Interferometry is the only chance to go further in this study. In this respect, 
ALMA will enable considerable progress.  

\begin{acknowledgements}
B. Parise wishes to thank the IRAM 30\,m operator Manuel Ruiz for his precious help 
during the observations. 

\end{acknowledgements}

\bibliographystyle{aa}
\bibliography{/Users/bparise/These/Manuscrit/biblio}

\begin{thebibliography}{38}
\expandafter\ifx\csname natexlab\endcsname\relax\def\natexlab#1{#1}\fi

\bibitem[{{Andr\'e} {et~al.}(2000){Andr\'e}, {Ward-Thompson}, \&
  {Barsony}}]{Andre00}
{Andr\'e}, P., {Ward-Thompson}, D., \& {Barsony}, M. 2000, Protostars and
  Planets IV, 59

\bibitem[{Bacmann {et~al.}(2002)Bacmann, Lefloch, Ceccarelli, Castets,
  Steinacker, \& Loinard}]{Bacmann02}
Bacmann, A., Lefloch, B., Ceccarelli, C., {et~al.} 2002, A\&A, 389, L6

\bibitem[{{Bacmann} {et~al.}(2003){Bacmann}, {Lefloch}, {Ceccarelli},
  {Steinacker}, {Castets}, \& {Loinard}}]{Bacmann03}
{Bacmann}, A., {Lefloch}, B., {Ceccarelli}, C., {et~al.} 2003, \apjl, 585, L55

\bibitem[{{Bottinelli} {et~al.}(2004){Bottinelli}, {Ceccarelli}, {Lefloch},
  {Williams}, {Castets}, {Caux}, {Cazaux}, {Maret}, {Parise}, \&
  {Tielens}}]{Bottinelli04a}
{Bottinelli}, S., {Ceccarelli}, C., {Lefloch}, B., {et~al.} 2004, \apj, 615,
  354

\bibitem[{{Caselli} {et~al.}(2003){Caselli}, {van der Tak}, {Ceccarelli}, \&
  {Bacmann}}]{Caselli03}
{Caselli}, P., {van der Tak}, F.~F.~S., {Ceccarelli}, C., \& {Bacmann}, A.
  2003, \aap, 403, L37

\bibitem[{{Cazaux} {et~al.}(2003){Cazaux}, {Tielens}, {Ceccarelli}, {Castets},
  {Wakelam}, {Caux}, {Parise}, \& {Teyssier}}]{Cazaux03}
{Cazaux}, S., {Tielens}, A.~G.~G.~M., {Ceccarelli}, C., {et~al.} 2003, \apjl,
  593, L51

\bibitem[{Ceccarelli {et~al.}(1998)Ceccarelli, Castets, Loinard, Caux, \&
  Tielens}]{Ceccarelli98}
Ceccarelli, C., Castets, A., Loinard, L., Caux, E., \& Tielens, A. G. G.~M.
  1998, A\&A, 338, L43

\bibitem[{Ceccarelli {et~al.}(2001)Ceccarelli, Loinard, Castets, Tielens, Caux,
  Lefloch, \& Vastel}]{Ceccarelli01}
Ceccarelli, C., Loinard, L., Castets, A., {et~al.} 2001, A\&A, 372, 998

\bibitem[{{Cernis}(1990)}]{Cernis90}
{Cernis}, K. 1990, \apss, 166, 315

\bibitem[{Charnley {et~al.}(1997)Charnley, Tielens, \& Rodgers}]{Charnley97}
Charnley, S.~B., Tielens, A. G. G.~M., \& Rodgers, S.~D. 1997, ApJ, 482, L203

\bibitem[{{Dartois} {et~al.}(1999){Dartois}, {Schutte}, {Geballe}, {Demyk},
  {Ehrenfreund}, \& {D'Hendecourt}}]{Dartois99}
{Dartois}, E., {Schutte}, W., {Geballe}, T.~R., {et~al.} 1999, \aap, 342, L32

\bibitem[{{Dartois} {et~al.}(2003){Dartois}, {Thi}, {Geballe}, {Deboffle},
  {d'Hendecourt}, \& {van Dishoeck}}]{Dartois03}
{Dartois}, E., {Thi}, W.-F., {Geballe}, T.~R., {et~al.} 2003, \aap, 399, 1009

\bibitem[{{Frerking} {et~al.}(1982){Frerking}, {Langer}, \&
  {Wilson}}]{Frerking82}
{Frerking}, M.~A., {Langer}, W.~D., \& {Wilson}, R.~W. 1982, \apj, 262, 590

\bibitem[{{Geppert} \& {al.}(2006)}]{Geppert06}
{Geppert}, W. \& {al.} 2006, in Astrochemistry, Recent successes and current
  challenges, ed. D.C. Lis, in press

\bibitem[{{Herbst}(2005)}]{Herbst05}
{Herbst}, E. 2005, in The Dusty and Molecular Universe: A Prelude to Herschel
  and ALMA, 205--210

\bibitem[{{J{\o}rgensen} {et~al.}(2002){J{\o}rgensen}, {Sch{\" o}ier}, \& {van
  Dishoeck}}]{Jorgensen02}
{J{\o}rgensen}, J.~K., {Sch{\" o}ier}, F.~L., \& {van Dishoeck}, E.~F. 2002,
  \aap, 389, 908

\bibitem[{{J{\o}rgensen} {et~al.}(2005){J{\o}rgensen}, {Sch{\" o}ier}, \& {van
  Dishoeck}}]{Jorgensen05}
{J{\o}rgensen}, J.~K., {Sch{\" o}ier}, F.~L., \& {van Dishoeck}, E.~F. 2005,
  \aap, 437, 501

\bibitem[{Loinard {et~al.}(2001)Loinard, Castets, Ceccarelli, Caux, \&
  Tielens}]{Loinard01}
Loinard, L., Castets, A., Ceccarelli, C., Caux, E., \& Tielens, A. G. G.~M.
  2001, ApJ, 552, 163

\bibitem[{{Loinard} {et~al.}(2002){Loinard}, {Castets}, {Ceccarelli},
  {Lefloch}, {Benayoun}, {Caux}, {Vastel}, {Dartois}, \& {Tielens}}]{Loinard02}
{Loinard}, L., {Castets}, A., {Ceccarelli}, C., {et~al.} 2002, \planss, 50,
  1205

\bibitem[{Loinard {et~al.}(2000)Loinard, Castets, Ceccarelli, Tielens, Faure,
  Caux, \& Duvert}]{Loinard00}
Loinard, L., Castets, A., Ceccarelli, C., {et~al.} 2000, A\&A, 359, 1169

\bibitem[{{Maret} {et~al.}(2004){Maret}, {Ceccarelli}, {Caux}, {Tielens},
  {J{\o}rgensen}, {van Dishoeck}, {Bacmann}, {Castets}, {Lefloch}, {Loinard},
  {Parise}, \& {Sch{\" o}ier}}]{Maret04}
{Maret}, S., {Ceccarelli}, C., {Caux}, E., {et~al.} 2004, \aap, 416, 577

\bibitem[{{Maret} {et~al.}(2005){Maret}, {Ceccarelli}, {Tielens}, {Caux},
  {Lefloch}, {Faure}, {Castets}, \& {Flower}}]{Maret05}
{Maret}, S., {Ceccarelli}, C., {Tielens}, A.~G.~G.~M., {et~al.} 2005, \aap,
  442, 527

\bibitem[{{Menten} {et~al.}(1988){Menten}, {Walmsley}, {Henkel}, \&
  {Wilson}}]{Menten88}
{Menten}, K.~M., {Walmsley}, C.~M., {Henkel}, C., \& {Wilson}, T.~L. 1988,
  \aap, 198, 253

\bibitem[{{Nagaoka} {et~al.}(2005){Nagaoka}, {Watanabe}, \&
  {Kouchi}}]{Nagaoka05}
{Nagaoka}, A., {Watanabe}, N., \& {Kouchi}, A. 2005, \apjl, 624, L29

\bibitem[{{Osamura} {et~al.}(2004){Osamura}, {Roberts}, \&
  {Herbst}}]{Osamura04}
{Osamura}, Y., {Roberts}, H., \& {Herbst}, E. 2004, \aap, 421, 1101

\bibitem[{{Osamura} {et~al.}(2005){Osamura}, {Roberts}, \&
  {Herbst}}]{Osamura05}
{Osamura}, Y., {Roberts}, H., \& {Herbst}, E. 2005, \apj, 621, 348

\bibitem[{{Parise} {et~al.}(2004){Parise}, {Castets}, {Herbst}, {Caux},
  {Ceccarelli}, {Mukhopadhyay}, \& {Tielens}}]{Parise04}
{Parise}, B., {Castets}, A., {Herbst}, E., {et~al.} 2004, \aap, 416, 159

\bibitem[{{Parise} {et~al.}(2005){Parise}, {Caux}, {Castets}, {Ceccarelli},
  {Loinard}, {Tielens}, {Bacmann}, {Cazaux}, {Comito}, {Helmich}, {Kahane},
  {Schilke}, {van Dishoeck}, {Wakelam}, \& {Walters}}]{Parise05a}
{Parise}, B., {Caux}, E., {Castets}, A., {et~al.} 2005, \aap, 431, 547

\bibitem[{{Parise} {et~al.}(2002){Parise}, {Ceccarelli}, {Tielens}, {Herbst},
  {Lefloch}, {Caux}, {Castets}, {Mukhopadhyay}, {Pagani}, \&
  {Loinard}}]{Parise02}
{Parise}, B., {Ceccarelli}, C., {Tielens}, A.~G.~G.~M., {et~al.} 2002, \aap,
  393, L49

\bibitem[{{Parise} {et~al.}(2003){Parise}, {Simon}, {Caux}, {Dartois},
  {Ceccarelli}, {Rayner}, \& {Tielens}}]{Parise03}
{Parise}, B., {Simon}, T., {Caux}, E., {et~al.} 2003, \aap, 410, 897

\bibitem[{{Roberts} {et~al.}(2002){Roberts}, {Fuller}, {Millar}, {Hatchell}, \&
  {Buckle}}]{Roberts02a}
{Roberts}, H., {Fuller}, G.~A., {Millar}, T.~J., {Hatchell}, J., \& {Buckle},
  J.~V. 2002, \aap, 381, 1026

\bibitem[{{Roberts} {et~al.}(2003){Roberts}, {Herbst}, \& {Millar}}]{Roberts03}
{Roberts}, H., {Herbst}, E., \& {Millar}, T.~J. 2003, \apjl, 591, L41

\bibitem[{{Roberts} {et~al.}(2004){Roberts}, {Herbst}, \& {Millar}}]{Roberts04}
{Roberts}, H., {Herbst}, E., \& {Millar}, T.~J. 2004, \aap, 424, 905

\bibitem[{Roberts \& Millar(2000)}]{Roberts00b}
Roberts, H. \& Millar, T.~J. 2000, A\&A, 364, 780

\bibitem[{{Stantcheva} \& {Herbst}(2003)}]{Stantcheva03}
{Stantcheva}, T. \& {Herbst}, E. 2003, \mnras, 340, 983

\bibitem[{{Tafalla} {et~al.}(2002){Tafalla}, {Myers}, {Caselli}, {Walmsley}, \&
  {Comito}}]{Tafalla02}
{Tafalla}, M., {Myers}, P.~C., {Caselli}, P., {Walmsley}, C.~M., \& {Comito},
  C. 2002, \apj, 569, 815

\bibitem[{Tielens(1983)}]{Tielens83}
Tielens, A. G. G.~M. 1983, A\&A, 119, 177

\bibitem[{{Vastel} {et~al.}(2004){Vastel}, {Phillips}, \& {Yoshida}}]{Vastel04}
{Vastel}, C., {Phillips}, T.~G., \& {Yoshida}, H. 2004, \apjl, 606, L127

\end{thebibliography}

\section*{Appendix : Computation of error bars for the fractionation ratios}

The ratio of two Gaussian distributed variables X and Y is Gaussian only when ${\rm \sigma_X \ll \bar{X}}$ and  ${\rm \sigma_Y \ll \bar{Y}}$. When the errors are too large, deviations from Gaussianity are observed and it is thus necessary  to compute accurately the repartition function of the random variable R=X/Y. This repartition function is given by:
$${\rm  f_{R} (u) = \int_{-\infty}^{\infty} f_{X}(uy)~f_{Y}(y)~|y|~dy } $$
This distribution is generally asymmetric, and thus leads to asymmetric confidence intervals. We chose in this paper to give intervals associated to a 90\% confidence level.
As an example, we present here the comparison with the repartition function for the CH$_2$DOH/CH$_3$OH ratio in IRAS2, assuming that CH$_2$DOH and CH$_3$OH have a Gaussian distribution.

\begin{figure}[!htbp]
\begin{center}
\includegraphics[width=0.5\textwidth]{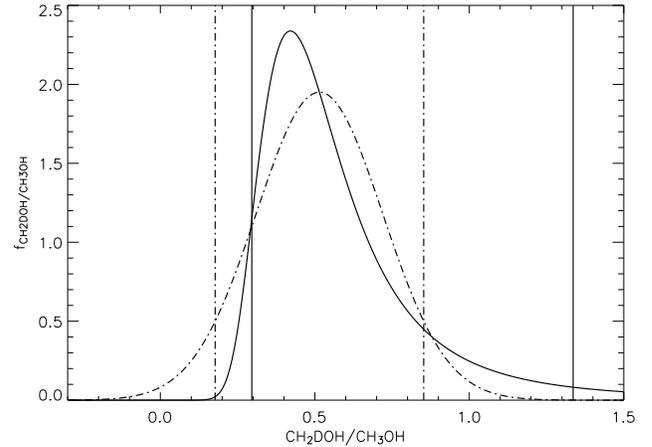}
\caption{Repartition function for the CH$_2$DOH/CH$_3$OH ratio (solid line). Overlaid in dash-dot   line is the repartition function computed assuming that the ratio is Gaussian, with a width computed by the standard error propagation. Vertical lines delimitate the 90\% confidence intervals for the two distributions. }
\label{ratio}
\end{center}
\end{figure}

\bigskip
\clearpage

\section*{ONLINE MATERIAL}

\begin{table*}[!ht]
\caption{Main-beam intensities, peak temperatures and width for the observed transitions towards NGC1333-IRAS4A, IRAS4B and IRAS2. The errors on the fluxes were computed as the quadratic sum of the statistical error and the calibration uncertainty (see text). Upper limits are 3$\sigma$ (see text).}
\label{flux}
\begin{tabular}{lcccccccccc}
\noalign{\smallskip}
\hline
\hline
\noalign{\smallskip}
 & & \multicolumn{3}{c}{IRAS4A} & \multicolumn{3}{c}{IRAS4B} &\multicolumn{3}{c}{IRAS2}\\ 
Freq. & E$_{up}$ & $\int{{\rm T}_{\rm mb}{\rm dv}}$ & T$_{\rm mb}$ &  $\Delta$v & $\int{{\rm T}_{\rm mb}{\rm dv}}$ & T$_{\rm mb}$ &  $\Delta$v & $\int{{\rm T}_{\rm mb}{\rm dv}}$ & T$_{\rm mb}$ &  $\Delta$v \\
GHz & cm$^{-1}$ & K.km.s$^{-1}$ & K & km.s$^{-1}$& K.km.s$^{-1}$ & K & km.s$^{-1}$& K.km.s$^{-1}$ & K & km.s$^{-1}$\\
\noalign{\smallskip}
\hline
\noalign{\smallskip}
\multicolumn{11}{l}{\small HDCO} \\
\noalign{\smallskip}
\hline
\noalign{\smallskip}
134.285 & 12.3 &0.95$\pm$0.10 & 0.58 & 1.5$\pm$0.1 & 0.59$\pm$0.06 & 0.43 & 1.3$\pm$0.1 & 0.43$\pm$0.04 & 0.33 & 1.2$\pm$0.1 \\
201.341 & 19.0 &1.09$\pm$0.11 & 0.73 & 1.5$\pm$0.1 & 0.70$\pm$0.07 & 0.47 & 1.4$\pm$0.1 & 0.69$\pm$0.07 & 0.53 & 1.2$\pm$0.1 \\
246.924 &  26.1 &0.98$\pm$0.11 & 0.65 & 1.6$\pm$0.1 & 0.74$\pm$0.10 & 0.57 & 1.2$\pm$0.1 & 0.50$\pm$0.08 & 0.33 & 1.4$\pm$0.2 \\
256.585 & 21.4 &1.19$\pm$0.13 & 0.94 & 1.2$\pm$0.1 & 1.38$\pm$0.15 & 0.83 & 1.6$\pm$0.1 & 1.11$\pm$0.14 & 0.92 & 1.1$\pm$0.1 \\
268.292 & 27.9 &0.82$\pm$0.13 & 0.64 & 1.2$\pm$0.1 & 0.74$\pm$0.12 & 0.44 & 1.6$\pm$0.1 & 1.00$\pm$0.16 & 0.76 & 1.1$\pm$0.1 \\
\noalign{\smallskip}
\hline
\noalign{\smallskip}
\multicolumn{11}{l}{\small D$_2$CO} \\
\noalign{\smallskip}
\hline
\noalign{\smallskip}
110.838 & 9.3 &0.16$\pm$0.01 & 0.12 & 1.2$\pm$0.1 & 0.095$\pm$0.011 & 0.12 & 0.8$\pm$0.1 & 0.077$\pm$0.017 & 0.062 & 1.2$\pm$0.4 \\
166.103 & 14.8 &0.33$\pm$0.04 & 0.24 & 1.3$\pm$0.2 & 0.24$\pm$0.04 & 0.22 & 1.0$\pm$0.2 & 0.095$\pm$0.022 & 0.21 & 0.4$\pm$0.1 \\
231.410 & 19.4 &0.37$\pm$0.05 & 0.28 & 1.3$\pm$0.1 & 0.47$\pm$0.06 & 0.41 & 1.1$\pm$0.1 & 0.63$\pm$0.07 & 0.76 & 0.8$\pm$0.1 \\
275.671 & 20.9 &--- & --- & --- & $\le$ 0.16 & - & - & $\le$ 0.26 & - & - \\
276.060 & 31.4 &--- & --- & --- & 0.10$\pm$0.03 & 0.46 & 0.2$\pm$0.1 & 0.44$\pm$0.10 & 0.45 & 0.9$\pm$0.2 \\
\noalign{\smallskip}
\hline
\noalign{\smallskip}
\multicolumn{11}{l}{\small CH$_2$DOH} \\
\noalign{\smallskip}
\hline
\noalign{\smallskip}
89.251 & 17.2 & 0.042$\pm$0.008 & 0.014 & 2.8$\pm$0.6 & 0.011$\pm$0.005 & 0.008& 1.3$\pm$0.8 & 0.021$\pm$0.008 & 0.006 & 3.4$\pm$1.2 \\
89.275 & 13.8 & 0.033$\pm$0.007 & 0.015 & 2.0$\pm$0.5 & 0.023$\pm$0.004 & 0.015& 1.4$\pm$0.3 & $\le$ 0.013      & -     & -           \\ 
89.408 & 4.5 &0.098$\pm$0.008 & 0.054 & 1.7$\pm$0.1 & 0.113$\pm$0.008 & 0.072& 1.5$\pm$0.1 & 0.038$\pm$0.005 & 0.015 & 2.3$\pm$0.4 \\ 
110.105 & 83.4 & 0.054$\pm$0.014 & 0.022 & 2.3$\pm$0.6 & $\le$ 0.027 & - & -                  & 0.050$\pm$0.009 & 0.025 & 1.8$\pm$3.4 \\
133.847 & 18.3 &0.084$\pm$0.022 & 0.046 & 1.7$\pm$0.4 & $\le$ 0.077 & - & -                  & 0.090$\pm$0.017 & 0.045 & 1.9$\pm$0.3 \\
133.873 &  21.7 &0.064$\pm$0.019 & 0.025 & 2.4$\pm$0.7 & 0.035$\pm$0.009 & 0.030 & 1.1$\pm$0.2 & 0.104$\pm$0.020 & 0.034 & 2.9$\pm$0.5 \\
133.882 & 33.6 & 0.054$\pm$0.019 & 0.018 & 2.6$\pm$0.9 & $\le$ 0.039      & -     & -           & 0.048$\pm$0.026 & 0.016 & 2.9$\pm$2.0 \\
133.893 & 27.4 & $\le$ 0.052      & -     & -           & 0.055$\pm$0.018 & 0.024 & 2.1$\pm$0.7 & $\le$ 0.10      &   -   &        -    \\
133.930 & 27.4& $\le$ 0.052      & -     & -           & $\le$0.039      & -     & -           & 0.080$\pm$0.022 & 0.025 & 3.0$\pm$0.7 \\
134.066 & 8.9 & 0.157$\pm$0.022 & 0.107 & 1.4$\pm$0.2 & 0.197$\pm$0.024 & 0.160 & 1.2$\pm$0.1 & 0.115$\pm$0.021 & 0.042 & 2.6$\pm$0.4 \\
134.112 & 20.2 & 0.068$\pm$0.015 & 0.071 & 0.9$\pm$0.2 & $\le$ 0.052 & - & - & $\le$ 0.065 & - & - \\
207.781 & 15.9 & 0.094$\pm$0.017 & 0.067 & 1.3$\pm$0.2 & 0.089$\pm$0.017 & 0.024& 3.6$\pm$0.6 & 0.144$\pm$0.029& 0.067 & 2.0$\pm$0.5 \\
223.071 & 33.6 & 0.116$\pm$0.022 & 0.059 & 1.9$\pm$0.3 & 0.042$\pm$0.011 & 0.034& 1.1$\pm$0.3 & 0.229$\pm$0.035& 0.081 & 2.7$\pm$0.4 \\
223.107 & 35.1& $\le$ 0.10      & -     & -           & 0.030$\pm$0.010 & 0.024& 1.2$\pm$0.4 & 0.234$\pm$0.041& 0.098 & 2.3$\pm$0.4\\
223.128 & 40.8 & $\le$ 0.083 & - & - & 0.046$\pm$0.013 & 0.037& 1.2$\pm$0.4 & 0.189$\pm$0.055& 0.073 & 2.4$\pm$0.8 \\
223.131 &  79.4 & 0.015$\pm$0.01 & 0.025 & 0.6$\pm$0.2  & 0.02$\pm$0.01 & 0.02 &0.9$\pm$0.3 & 0.04$\pm$0.03& 0.016& 2.3$\pm$1.2 \\
223.131 & 79.4 & 0.015$\pm$0.01 & 0.025 & 0.6$\pm$0.2  & 0.02$\pm$0.01 & 0.02 &0.9$\pm$0.3 & 0.04$\pm$0.03& 0.016& 2.3$\pm$1.2 \\
223.154 & 60.8 & 0.046$\pm$0.02 & 0.02 & 2.0$\pm$0.5  & 0.02$\pm$0.01 & 0.015 &1.2$\pm$0.5 & 0.13$\pm$0.03 & 0.05 & 2.6$\pm$0.4 \\
223.154 & 60.8 & 0.046$\pm$0.02 & 0.02 & 2.0$\pm$0.5  & 0.02$\pm$0.01 & 0.015 &1.2$\pm$0.5 & 0.13$\pm$0.03 & 0.05 & 2.6$\pm$0.4 \\
223.315 & 40.8 & 0.061$\pm$0.016& 0.040& 1.4$\pm$0.4 & 0.083$\pm$0.015 & 0.032 &2.5$\pm$0.4 & 0.192$\pm$0.032& 0.061 & 3.0$\pm$0.4 \\
223.422 & 33.6 & 0.126$\pm$0.024& 0.061 & 2.0$\pm$0.4 & 0.062$\pm$0.010& 0.042 & 1.4$\pm$0.2 & 0.229$\pm$0.038& 0.064 & 3.4$\pm$0.5 \\
\noalign{\smallskip}
\hline
\noalign{\smallskip}
\multicolumn{11}{l}{\small CH$_3$OD}   \\
\noalign{\smallskip}
\hline
\noalign{\smallskip}
110.189 & 7.8 & $\le$ 0.03 &  - &  - & 0.022$\pm$0.009  & 0.054  &  0.5$\pm$0.2 &  $\le$ 0.025 &  -  &  -  \\
110.263 & 10.8 & 0.041$\pm$0.014 & 0.018  & 2.0$\pm$0.7 & $\le$ 0.03 & - & - & $\le$ 0.02 &  - &  -  \\
110.476 & 15.4 & $\le$ 0.02 &  - &  - & $\le$ 0.02 &  - &  - & $\le$ 0.02 &  - &  - \\
133.925 & 6.0 & 0.076$\pm$0.020 & 0.024 & 2.7$\pm$0.8 &  0.046$\pm$0.015 & 0.038 &  1.1$\pm$0.5 & $\le$ 0.05 &  - & - \\
223.309 & 26.8 & $\le$ 0.07 & - & - & 0.035$\pm$0.009 & 0.029 & 1.2$\pm$0.3 & $\le$ 0.15 & - & -   \\
226.539 & 22.7 & $\le$ 0.06 & - &  - & $\le$ 0.04 & - &  - & $\le$ 0.17 &  - &  - \\
\noalign{\smallskip}
\hline
\noalign{\smallskip}
\multicolumn{11}{l}{\small CHD$_2$OH}   \\
\noalign{\smallskip}
\hline
\noalign{\smallskip}
83.1292 & 17.0 & $\le$ 0.03 & - & - & 0.018$\pm$0.005 & 0.026 & 0.7$\pm$0.2 & $\le$ 0.021 & - & - \\
83.2895 & 4.2 & 0.009$\pm$0.002 & 0.016 & 0.5$\pm$0.1 & 0.024$\pm$0.004 & 0.027 & 0.9$\pm$0.2 & $\le$ 0.021 &  - & - \\
83.3036 & 10.3 & 0.015$\pm$0.006 & - & - & 0.011$\pm$0.004 & 0.012 & 0.9$\pm$0.3 &  $\le$ 0.021  &  - &  - \\ 
207.771 & 33.6 & 0.066$\pm$0.040 & 0.022 & 2.8$\pm$1.7 & 0.089$\pm$0.025 & 0.025 & 3.4$\pm$0.9 & 0.115$\pm$0.041 & 0.029 & 3.7$\pm$1.4 \\
207.827 & 42.6 & $\le$ 0.07 & - & - & 0.097$\pm$0.027 & 0.021 & 4.4$\pm$1.0 & 0.087$\pm$0.045 & 0.016 & 5.2$\pm$2.3 \\
207.864 & 68.4 & 0.062$\pm$0.030 & 0.036 & 1.6$\pm$0.7 & $\le$ 0.035 & - & - & 0.088$\pm$0.033 & 0.035 & 2.3$\pm$1.0 \\
207.868 & 53.5 & 0.12$\pm$0.04 & 0.03 & 3.7$\pm$0.8 & 0.034$\pm$0.011 & 0.02 & 1.8$\pm$0.4 & 0.13$\pm$0.03 & 0.03 & 4.8$\pm$0.9 \\
207.869 & 53.5 & 0.12$\pm$0.04 & 0.03 & 3.7$\pm$0.8 & 0.034$\pm$0.011 & 0.02 & 1.8$\pm$0.4 & 0.13$\pm$0.03 & 0.03 & 4.8$\pm$0.9 \\
\noalign{\smallskip}
\hline
\noalign{\smallskip}
\end{tabular}
\end{table*}

\begin{table*}[!htbp]
\caption{Main-beam intensities, peak temperatures and width for the detected transitions towards L1448N, L1448mm and L1157mm. The errors on the fluxes were computed as the quadratic sum of the statistical error and the calibration uncertainty (see text). Upper limits are 3$\sigma$. Upper limits for lines not detected in any of these three sources can be found in Tab. \ref{flux2}. }
\label{flux1.5}
\begin{tabular}{lcccccccccc}
\noalign{\smallskip}
\hline
\hline
\noalign{\smallskip}
 & & \multicolumn{3}{c}{L1448N} & \multicolumn{3}{c}{L1448mm} &\multicolumn{3}{c}{L1157mm}\\ 
Freq. & E$_{up}$ & $\int{{\rm T}_{\rm mb}{\rm dv}}$ & T$_{\rm mb}$ &  $\Delta$v & $\int{{\rm T}_{\rm mb}{\rm dv}}$ & T$_{\rm mb}$ &  $\Delta$v & $\int{{\rm T}_{\rm mb}{\rm dv}}$ & T$_{\rm mb}$ &  $\Delta$v \\
GHz & cm$^{-1}$ & K.km.s$^{-1}$ & K & km.s$^{-1}$& K.km.s$^{-1}$ & K & km.s$^{-1}$& K.km.s$^{-1}$ & K & km.s$^{-1}$\\
\noalign{\smallskip}
\hline
\noalign{\smallskip}
\multicolumn{11}{l}{\small HDCO} \\
\noalign{\smallskip}
\hline
\noalign{\smallskip}
134.285 & 12.3 & 0.32$\pm$0.04 & 0.26 & 1.1$\pm$0.1 & 0.38$\pm$0.06 & 0.38 & 0.9$\pm$0.1 & 0.11$\pm$0.01 & 0.13 & 0.8$\pm$0.1 \\
201.341 & 19.0 & 0.36$\pm$0.05 & 0.38 & 1.0$\pm$0.1 & 0.53$\pm$0.06 & 0.97 & 0.6$\pm$0.1 & 0.10$\pm$0.03 & 0.12 & 0.8$\pm$0.1 \\
246.924 & 26.1 & 0.35$\pm$0.04 & 0.30 & 1.1$\pm$0.1 & 0.61$\pm$0.08 & 0.63 & 0.9$\pm$0.1 & 0.26$\pm$0.03 & 0.20 & 1.2$\pm$0.1 \\
256.585 & 21.4 & 0.60$\pm$0.10 & 0.47 & 1.2$\pm$0.2 & 0.79$\pm$0.10 & 1.11 & 0.7$\pm$0.1 & 0.18$\pm$0.03 & 0.18 & 0.9$\pm$0.1 \\
268.292 & 27.9 & 0.34$\pm$0.08 & 0.24 & 1.4$\pm$0.3 & 0.40$\pm$0.08 & 0.66 & 0.6$\pm$0.1 & 0.15$\pm$0.05 & 0.20 & 0.7$\pm$0.4 \\
\noalign{\smallskip}
\hline
\noalign{\smallskip}
\multicolumn{11}{l}{\small D$_2$CO} \\
\noalign{\smallskip}
\hline
\noalign{\smallskip}
110.838 & 9.3 & 0.074$\pm$0.010 & 0.137 & 0.5$\pm$0.1 & 0.113$\pm$0.014 & 0.10 & 1.1$\pm$0.1 & $\le$ 0.67 & - & - \\
166.103 & 14.8 & 0.088$\pm$0.018 & 0.204 & 0.4$\pm$0.1 & 0.114$\pm$0.021 & 0.27 & 0.4$\pm$0.1 & $\le$ 0.10 & - & - \\
231.410 & 19.4 & $\le$ 0.11      & -     & -           & 0.325$\pm$0.038 & 0.59 & 0.5$\pm$0.4 & 0.109$\pm$0.031 & 0.182 & 0.6$\pm$0.2 \\
275.671 & 20.9 & $\le$ 0.20      & -     & -           & $\le $0.16      & -    & -           & $\le$ 0.26 & - & - \\
276.060 & 31.4 & $\le$ 0.13      & -     & -           & 0.092$\pm$0.051 & 0.14 & 0.6$\pm$0.4 & $\le$ 0.21 & - & - \\
\noalign{\smallskip}
\hline
\noalign{\smallskip}
\multicolumn{11}{l}{\small CH$_2$DOH} \\
\noalign{\smallskip}
\hline
\noalign{\smallskip}
89.408 & & 0.029$\pm$0.002 & 0.053 & 0.5$\pm$0.1 & 0.040$\pm$0.007 & 0.034 & 1.1$\pm$0.2 & 0.021$\pm$0.004 & 0.057 & 0.3$\pm$1.0 \\
\noalign{\smallskip}
\hline
\noalign{\smallskip}
\multicolumn{11}{l}{\small CH$_3$OD}   \\
\noalign{\smallskip}
\hline
\noalign{\smallskip}
110.189 & & $\le$ 0.04 & - & - & 0.078$\pm$0.010 & 0.073 & 1.0$\pm$0.1 & 0.025$\pm$0.013 & 0.050 & 0.5$\pm$0.3 \\
\noalign{\smallskip}
\hline
\noalign{\smallskip}
\end{tabular}
\end{table*}

\begin{table}[!htbp]
\centering
\begin{tabular}{lcccc}
\noalign{\smallskip}
\hline
\hline
\noalign{\smallskip}
 & & \multicolumn{3}{c}{L1527} \\ 
Frequency & E$_{\rm up}$ &$\int{{\rm T}_{\rm mb}dv}$ & T$_{\rm mb}$ &  $\Delta$v \\
GHz & cm$^{-1}$ & K.km.s$^{-1}$ & K & km.s$^{-1}$\\
\noalign{\smallskip}
\hline
\noalign{\smallskip}
HDCO  &  & & & \\
\noalign{\smallskip}
\hline
\noalign{\smallskip}
134.285  &  12.3 & 0.75$\pm$0.08 & 1.1 & 0.7$\pm$0.1  \\ 
201.341  & 19.0 & 0.70$\pm$0.08 & 0.98 & 0.7$\pm$0.1  \\
256.585  & 21.4 & 0.77$\pm$0.10 & 0.18 & 0.9$\pm$0.1  \\
\noalign{\smallskip}
\hline
\noalign{\smallskip}
D$_2$CO & & & \\
\noalign{\smallskip}
\hline
\noalign{\smallskip}
110.838 &  9.3 & 0.133$\pm$0.016 & 0.243 & 0.5$\pm$0.1 \\
166.103 &  14.8 & 0.219$\pm$0.038 & 0.327 & 0.6$\pm$0.1 \\
231.410 &  19.4 & 0.367$\pm$0.047 & 0.601 & 0.6$\pm$0.1 \\
276.060 &  31.4 & $\le$\,0.23 & - & - \\
\noalign{\smallskip}
\hline
\noalign{\smallskip}
\end{tabular}
\caption{Main-beam intensities, peak temperatures and width for the observed transitions towards L1527.}
\label{L1527}
\end{table}

\begin{table}[!htbp]
\centering
\caption{Main-beam temperature upper limits at 3$\sigma$ level for the deuterated methanol transitions observed towards L1448N, L1448mm and L1157mm.} 
\label{flux2}
\begin{tabular}{lccc}
\noalign{\smallskip}
\hline
\hline
\noalign{\smallskip}
 & ~~L1448N~~ & ~~L1448mm~~ & ~~L1157mm~~\\ 
Freq. & $\int{{\rm T}_{\rm mb}{\rm dv}}$ &  $\int{{\rm T}_{\rm mb}{\rm dv}}$  & $\int{{\rm T}_{\rm mb}{\rm dv}}$  \\
GHz & K.km.s$^{-1}$ & K.km.s$^{-1}$ & K.km.s$^{-1}$ \\
\noalign{\smallskip}
\hline
\noalign{\smallskip}
\multicolumn{4}{l}{\small CH$_2$DOH} \\
\noalign{\smallskip}
\hline
\noalign{\smallskip}
89.251 & $\le$ 0.015 & $\le$ 0.03 & $\le$ 0.02 \\
89.275 & $\le$ 0.015 & $\le$ 0.03 & $\le$ 0.02 \\
110.105 & $\le$ 0.03 & $\le$ 0.05 & $\le$ 0.04 \\
207.781 & $\le$ 0.08 & $\le$ 0.08 & $\le$ 0.05 \\
223.071 & $\le$ 0.05 & $\le$ 0.08 & $\le$ 0.05 \\
223.107 & $\le$ 0.05 & $\le$ 0.08 & $\le$ 0.05 \\
223.128 & $\le$ 0.05 & $\le$ 0.08 & $\le$ 0.05 \\
223.131 & $\le$ 0.05 & $\le$ 0.08 & $\le$ 0.05 \\
223.131 & $\le$ 0.05 & $\le$ 0.08 & $\le$ 0.05 \\
223.154 & $\le$ 0.05 & $\le$ 0.08 & $\le$ 0.05 \\
223.154 & $\le$ 0.05 & $\le$ 0.08 & $\le$ 0.05 \\
223.315 & $\le$ 0.04 & $\le$ 0.06 & $\le$ 0.05 \\
223.422 & $\le$ 0.04 & $\le$ 0.06 & $\le$ 0.04 \\
\noalign{\smallskip}
\hline
\noalign{\smallskip}
\multicolumn{4}{l}{\small CH$_3$OD}   \\
\noalign{\smallskip}
\hline
\noalign{\smallskip}
110.263 & $\le$ 0.03 & $\le$ 0.05 & $\le$ 0.03 \\
110.476 & $\le$ 0.02 & $\le$ 0.04 & $\le$ 0.03 \\
223.309 & $\le$ 0.04 & $\le$ 0.06 & $\le$ 0.05 \\
\noalign{\smallskip}
\hline
\noalign{\smallskip}
\multicolumn{4}{l}{\small CHD$_2$OH}   \\
\noalign{\smallskip}
\hline
\noalign{\smallskip}
207.771 & $\le$ 0.08 & $\le$ 0.08 & $\le$ 0.04 \\
207.827 & $\le$ 0.08 & $\le$ 0.08 & $\le$ 0.04 \\
207.864 & $\le$ 0.08 & $\le$ 0.08 & $\le$ 0.04 \\
207.868 & $\le$ 0.08 & $\le$ 0.08 & $\le$ 0.04 \\
207.869 & $\le$ 0.08 & $\le$ 0.08 & $\le$ 0.04 \\
\noalign{\smallskip}
\hline
\noalign{\smallskip}
\end{tabular}
\end{table}

\end{document}